\documentclass[review]{elsarticle}

\usepackage{lineno,hyperref}
\usepackage[T1]{fontenc}
\usepackage{pdfpages}
\usepackage{etoolbox}

\modulolinenumbers[5]

\journal{Journal of \LaTeX\ Templates}

\begin{document}


\clearpage
\includepdf[pages={1}]{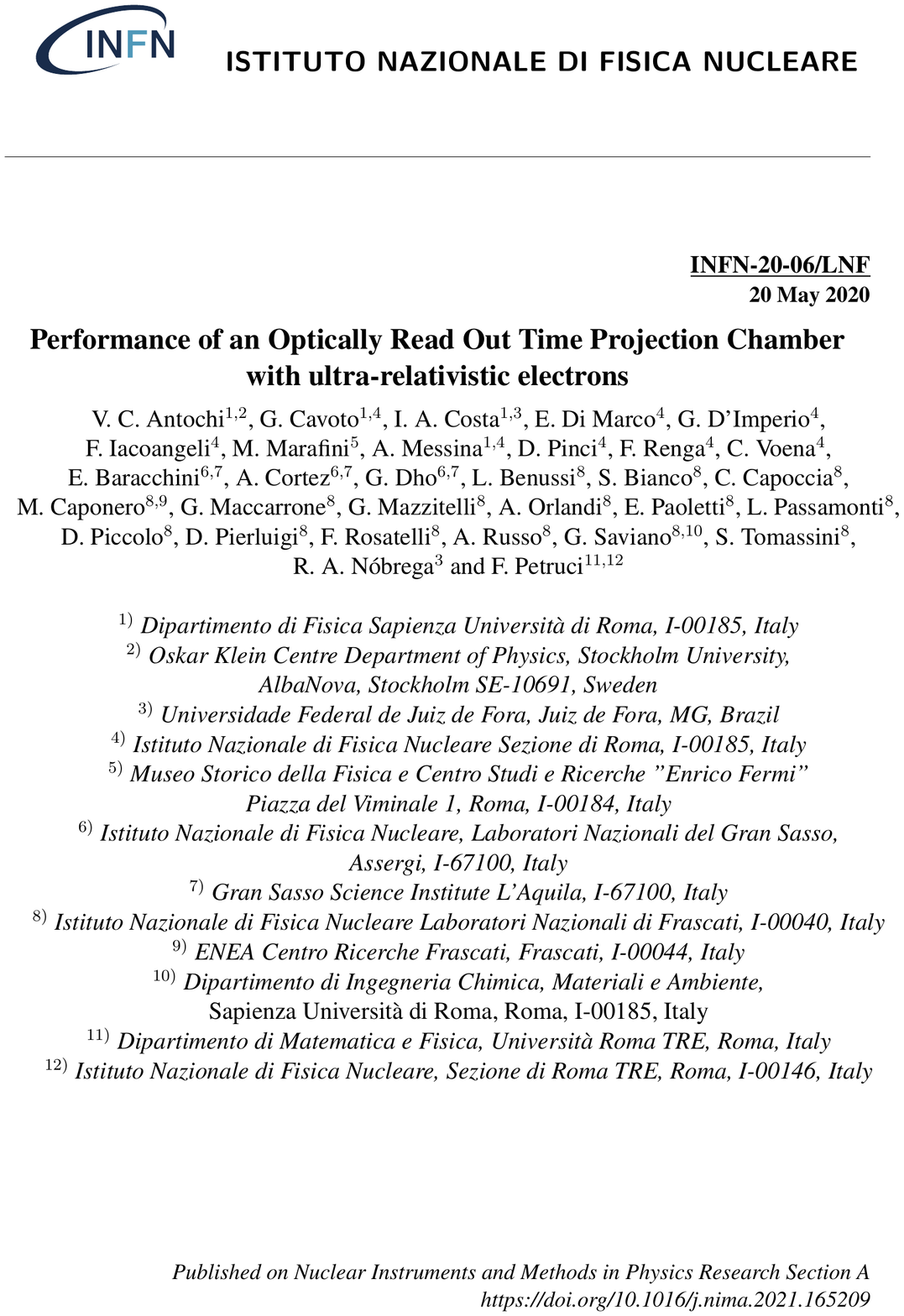}
\clearpage

\makeatletter
\patchcmd{\@outputpage@head}{\@ifx{\LS@rot\@undefined}{}{\LS@rot}}{}{}{}
\makeatother


\begin{frontmatter}

\title{Performance of an Optically Read Out Time Projection Chamber with ultra-relativistic electrons}

\author[a1,a2]{V. C. Antochi}
\author[a3]{E.~Baracchini}
\author[a4]{L.~Benussi}
\author[a4]{S.~Bianco}
\author[a4]{C.~Capoccia}
\author[a5,a4]{M.~Caponero}
\author[a1,a7]{G.~Cavoto}
\author[a3]{A.~Cortez}
\author[a1,a6]{I.~A.~Costa}
\author[a7]{E.~Di Marco}
\author[a7]{G.~D'Imperio}
\author[a3]{G.~Dho}
\author[a7]{F.~Iacoangeli}
\author[a4]{G~ Maccarrone}
\author[a8,a7]{M. Marafini}
\author[a4]{G.~Mazzitelli\corref{cor}}
\ead[cor]{giovanni.mazzitelli@lnf.infn.it}
\author[a1,a7]{A.~Messina}
\author[a6]{R.~A.~Nóbrega}
\author[a4]{A.~Orlandi}
\author[a4]{E.~Paoletti}
\author[a4]{L.~Passamonti}
\author[a9,a10]{F.~Petrucci}
\author[a4]{D.~Piccolo}
\author[a4]{D.~Pierluigi}
\author[a7]{D.~Pinci}
\author[a7]{F.~Renga}
\author[a4]{F.~Rosatelli}
\author[a4]{A.~Russo}
\author[a11,a4]{G.~Saviano}
\author[a4]{S~Tomassini}
\author[a7]{C.~Voena}

\address[a1]{Dipartimento di Fisica, Sapienza Universit\`a di Roma, Rome I-00185, Italy}
\address[a2]{now at Oskar Klein Centre, Department of Physics, Stockholm University, AlbaNova, Stockholm SE-10691, Sweden}
\address[a3]{Gran~Sasso~Science~Institute, L'Aquila I-67100, Italy}
\address[a4]{Istituto Nazionale di Fisica Nucleare, Laboratori Nazionali di Frascati, I-00040, Italy}
\address[a5]{ENEA Centro Ricerche Frascati, Frascati (Rome) I-00044, Italy}
\address[a6]{Universidade Federal de Juiz de Fora, Juiz de Fora 36000-000, MG, Brazil}
\address[a7]{Istituto~Nazionale~di~Fisica~Nucleare~Sezione di Roma, Rome I-00185, Italy}
\address[a8]{Museo Storico della Fisica e Centro Studi e Ricerche "Enrico Fermi" Piazza del Viminale 1, Rome I-00184, Italy}
\address[a9]{Istituto Nazionale di Fisica Nucleare, Sezione di Roma TRE, Rome I-00146, Italy}
\address[a10]{Dipartimento di Matematica e Fisica, Universit\`a Roma TRE, Rome I-00146, Italy}
\address[a11]{Dipartimento di Ingegneria Chimica, Materiali e Ambiente, Sapienza Universit\`a di Roma, Rome I-00185, Italy}

\cortext[cor]{Corresponding author}

\begin{abstract}
The Time Projection Chamber (TPC) is an ideal candidate to finely study the  charged   particle ionization in a gaseous medium. Large volume TPCs  can be readout with a suitable number of channels offering a complete 3D reconstruction of a charged particle track, that is the sequence of its energy releases in the TPC gas volume. Moreover, He-based TPCs  are very promising  to study keV energy particles as nuclear recoils, opening the possibility for directional searches  of Dark Matter (DM) and the study of Solar Neutrinos (SN). 

In this paper we report the analysis of the data acquired with a small TPC prototype (named LEMOn) built by the CYGNO collaboration that was exposed to a beam of 450~MeV electrons at the Beam Test Facility of National Laboratories of Frascati.
LEMOn is operated with a He-CF$_4$ mixture at atmospheric pressure and is based on a Gas Electron Multipliers amplification stage that  produces visible light collected by the high granularity and very good sensitivity of scientific CMOS camera. This type of readout - in conjunction with a fast light detection - allows a 3D reconstruction of the electrons tracks.
The electrons are leaving a trail of clusters of ionizations corresponding to a few keV energy release each. Their study leads to  predict  a keV energy threshold and  1-10 mm longitudinal and  0.1-0.3 mm transverse  position resolution (sigma) for nuclear recoils, very  promising  for the application of  optically read out TPC  to DM searches and SN measurements.

\end{abstract}

\begin{keyword}
TPC, MPGD, GEM, Optical Readout, Tracking Detector, Dark Matter
\end{keyword}

\end{frontmatter}


\section{Introduction}

Large Time Projection Chambers (TPC)  have been  employed  in various high energy physics and nuclear physics experiments \cite{bib:tpc1,bib:tpc2,bib:tpc3,bib:tpc4}. These detectors consist of a relatively large gas volume where ionization electrons are drifted by means of a suitable electric field toward the readout plane (usually on one or two sides). The measurement of cluster positions and drift times allows to reconstruct charged particle's trajectories with sub-mm spatial resolution. 

 TPCs have been also proposed for ultra-rare events searches as nuclear and electron recoils in  the directional search for   Dark Matter (DM) and  results has been recently reported on these searches  \cite{Battat:2016xxe,  BATTAT20151,Ikeda:2020mvr}. Different types of electron amplification and readout have been studied\cite{Battat:2016pap} based on a  multi-wire proportional chamber scheme (DRIFT \cite{BATTAT20151,ALNER2005173}), on micro-pattern gaseous detectors (MPGD) using pixels or strips and a custom electronics for a complete 3D readout (MIMAC\cite{Riffard:2016mgw,Sauzet:2020dut}, NEWAGE \cite{Hashimoto:2017hlz}, D3 \cite{bib:vahsen}). Also the detection of neutrinos  coming from the Sun (SN) with large TPC  was proposed \cite{Seguinot:1992zu, ARPESELLA1996333}.
  
An alternative readout technique (referred as {\it optical readout}) is based on the detection of the light emitted by the gas molecules during the avalanche processes triggered by the primary ionization electrons\cite{bib:Fraga,bib:Margato1,bib:Margato2}. In this approach, the use of suitable lenses allows the reconstruction on the optical plane of the projection of the particle trajectory on the amplification stage.

A first directional DM detector prototype was using an optical readout with an image intensified CCD~\cite{Buckland}.  The DMTPC~\cite{Deaconu:2017vam} collaboration was then proposing to use a CCD with a wire mesh amplification stage in a low pressure CF$_4$ gas. A combined readout of a PMT and CCD  was also attempted for a 3D reconstruction~\cite{BATTAT20146, bib:pmt1}. 
 The use of   Gas Electron Multipliers (GEM)~\cite{Sauli:1997qp} as MPGD  amplifiers  was then introduced to study  low energy electron and nuclear recoils in low pressure CF$_4$ with a CCD readout~\cite{bib:loomba55Fe}.
Moreover, this optical approach was tested and developed in past years  for other applications~\cite{bib:opto1,bib:opto2,BRUNBAUER201824}.

  The recent introduction of
  low noise and high granularity scientific CMOS (sCMOS) sensors represented an important step forward in terms of granularity and readout noise with respect to the use of CCD. The better performance reflects on a better space resolution and a higher sensitivity which allow to clearly detect and to reconstruct very low energy releases while working at atmospheric pressure \cite{bib:jinst_orange1, bib:nim_orange2}.

The MPGD optical readout can in fact  represent an opportunity for large TPCs for rare events searches. The optical coupling locates  the sensors out of the sensitive volume, reducing the interference with the  MPGD high
voltage operation and  the gas contamination.
Moreover, the use of suitable lenses enables to image   large surfaces to small sensors.

In last years, optical sensors market had an impressive development: sensors able to provide the required large granularity along with a very low noise level and high sensitivity to single photon counting are now available.
On the other hand, the readout frame rate well below the kHz value still represents a limiting factor for these devices. 

Many smaller scale prototypes have been built and tested so far showing promising results (NITEC~\cite{JINST:nitec},ORANGE~\cite{NIM:Marafinietal, bib:jinst_orange2}).

   In this paper we describe the  data acquired  with a  7 litres prototype (named LEMOn -  Long Elliptical MOdule) exposed to 450 MeV electrons at the Beam Test Facility (BTF)~\cite{bib:btf} of INFN National Laboratory of Frascati (LNF).
    LEMOn has   an optical readout of a triple-GEM structure  based on one sCMOS camera  to image  energy releases in  He-based gas mixtures  kept  at atmospheric pressure. This would therefore allow to host larger target mass for DM searches than previous low pressure TPC prototypes and experiments.  Moreover, working at atmospheric pressure brings technological advantages allowing simpler constructive structures and
    easier operations.
    
    BTF ultra-relativistic electrons   are  leaving along their path a straight trail of  clusters of few ionization electrons each.  We  evaluate the capability  to detect within the LEMOn drift region short sequences of ionization clusters  ({\it segments }  of the track)  and to reconstruct  their  positions. Each of these segments  corresponds to   small energy deposits of the order of one keV. Even if the topology of the track segments is different from that   typical of the nuclear or electron recoils of the same energy, with this method keV energy releases can be easily characterized. This would open the possibility to operate He-based gas mixtures TPC with a keV  energy threshold, to be sensitive to a DM particle with a mass in the range of GeV \cite{Pinci:2019hhw}.

    The use of a beam with a precise time reference allows to operate LEMOn as a standard TPC to evaluate the absolute longitudinal distance.
    However, in absence of a reference time for the events as in a DM search, the ionization electron drift time cannot be exploited to determine  their initial longitudinal position inside the detector. In this paper, we study also a method to measure the  longitudinal position in the drift volume based on the electron diffusion, as already proposed for an  electronic readout (see for example \cite{bib:lewis,FENG201735}). This is in fact very important to fully define a fiducial gas volume in a DM search, in a way to discard the  background events  generated by the detector components, in particular the cathode and the amplification system (as the GEMs)\cite{Battat:2015rna,Daw:2013waa}.

LEMOn was also tested with other  particles sources~\cite{bib:eps, bib:ieee17, bib:elba, Costa:2019tnu}  as neutron beams and  various radioactive sources.
A design of a  1 m$^3$ demonstrator (named CYGNO) to be constructed in  2021-2022 and to be hosted at the INFN  National Laboratory of Gran Sasso (LNGS) will rely on the LEMOn perfomance.
The main sizes of LEMOn (20~cm maximum drift path and almost 500~cm$^2$ GEMs) are in fact of the same order of magnitude of the typical sizes of readout modules foreseen for the  1 m$^3$  scale (50~cm maximum drift path and around 1000~cm$^2$ GEMs).

 In a later phase, a 30-100 m$^3$ detector is foreseen, as an element of a world distributed observatory for DM and SN within the CYGNUS international network\cite{baracchini2019cygno, Abe:2020bbf,CYGNUSweb}.

\section{LEMOn Prototype Design}

\begin{figure*}[!ht]
\centering
\includegraphics[scale=0.4]{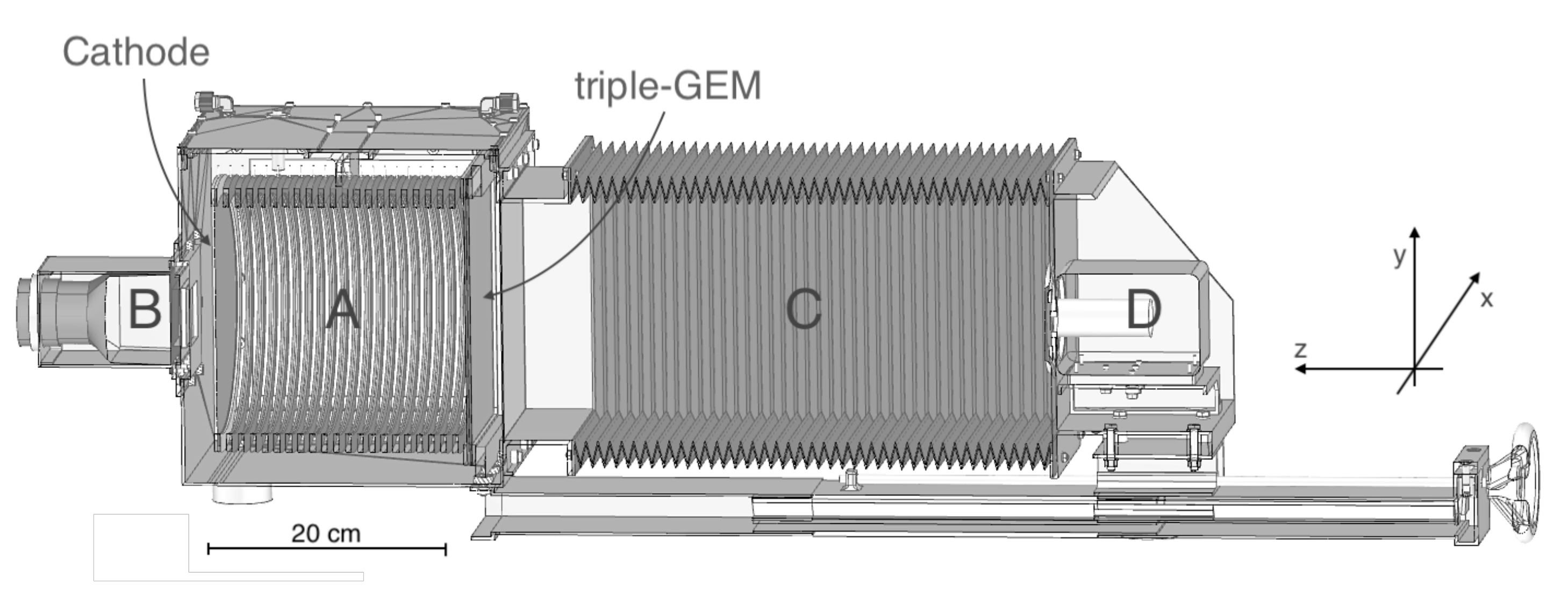}\DeclareGraphicsExtensions.
\caption{LEMOn design. PMT holder (B), semi-transparent cathode, field cage rings (A),  the triple GEM stack, the optical bellow (C) and the  ORCA Flash camera holder (D).
Reference frame is also indicated: the origin of the axis is located at the middle of the GEM plane in vertical Y coordinate, while the X origin is located at the beginning of the GEM plane to coincide with the particle track length; moreover, the Z coordinate represents the distance from the GEM plane starting from it.}
\label{fig:sex}
\end{figure*}

The LEMOn prototype structure (Fig.~\ref{fig:sex}) was made of  Acrylic Styrene Acrylonitrile  (ASA) at the 3D printing Facility of LNF\cite{3dprinting}. This has offered the opportunity  to easily design and to quickly develop detectors and also to test the 3D printing system for gas detector applications.
Obtained gas tightness was very satisfactory and the very good electrostatic stability of the prototype \cite{bib:stab} indicated a high level of cleanness in the sensitive and amplification volumes.

\begin{figure}[!ht]
\centering
\includegraphics[scale=0.290]{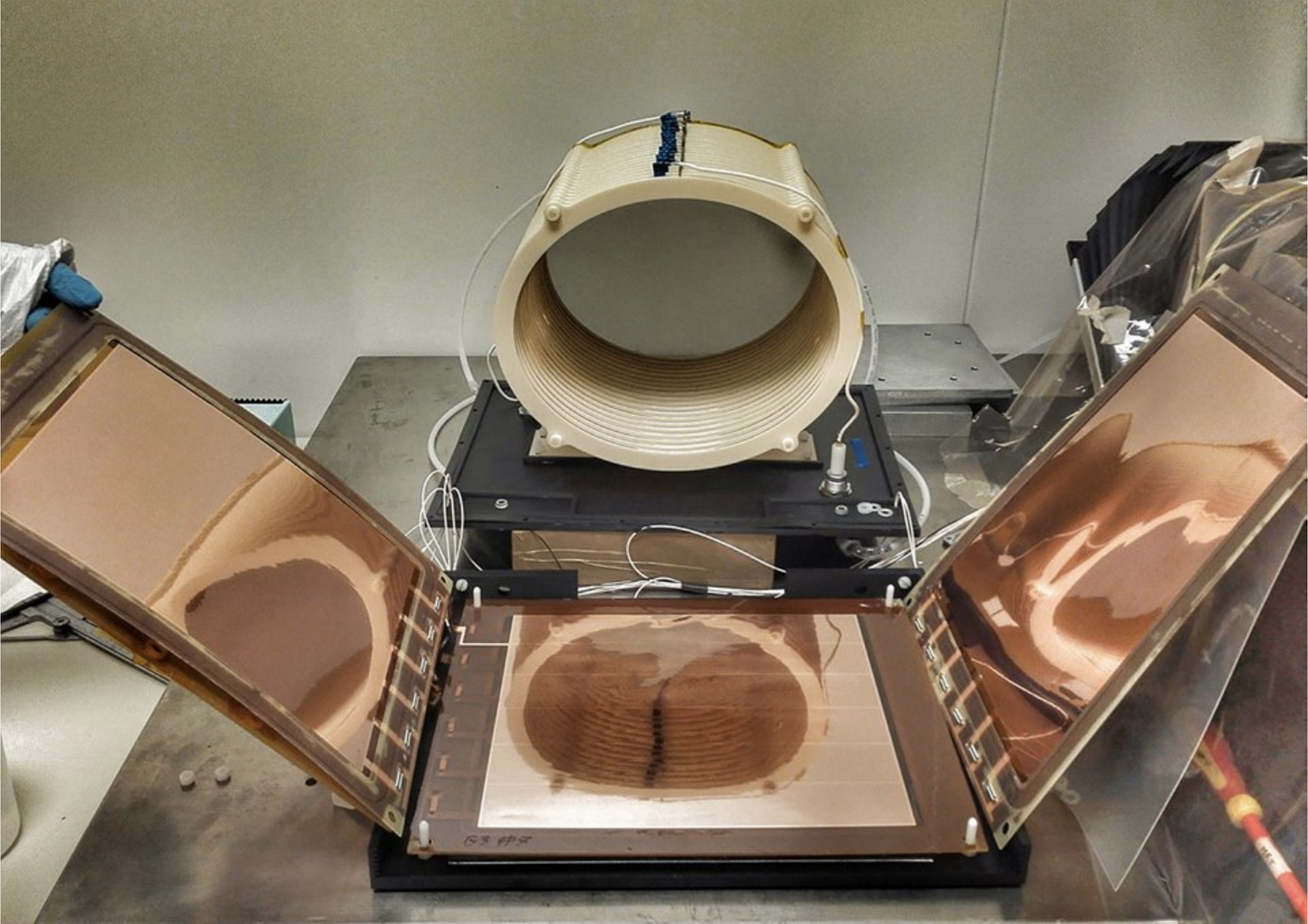}\DeclareGraphicsExtensions.
\caption{Exploded triple GEM structure (on the front)  with  the 3D-printed field cage rings (white)  equipped with a semi-transparency cathode (on the back of the rings).}
\label{fig:gem}
\end{figure}

The LEMOn's heart consists of a 7 liter active drift volume surrounded by a  200$\times$240~mm$^2$  field cage with a 200 mm distance between the anode and the cathode with an elliptical shape to avoid edge effects in the electric field intensity. The  field cage electrodes are   1~mm Cu+Ag wires held at their positions  by nineteen 3D-printed rings with 1~cm pitch. The anode side is instrumented with a 200$\times$240~mm$^2$ rectangular triple  GEM structure located at a position 10 mm apart from the last field cage ring. They are LHCb-like \cite{bib:thesis}   50~$\mu$m thick GEMs, with 70~$\mu$m diameter holes and 140~$\mu$m pitch   and  with two 2 mm wide  transfer  field gaps between them (Fig.~\ref{fig:gem}). Bottom electrode of last GEM was grounded and no anode was present below the last GEM.

A $203\times254\times1$ mm$^3$ transparent window 
and a  bellow with a tunable length allows  to collect the light emitted within the GEMs by using an ORCA-Flash 4.0 camera \cite{ORCAcamera} positioned  at a distance of about 50~cm from the outermost  GEM layer and based on an sCMOS sensor. The main feature of CMOS technology is that each individual pixel has its own amplifier. Therefore, pixel responses can be quickly processed, suppressing possible noise pick-up~\cite{bib:cmos}. Moreover, individual amplifiers allow to operate CMOS based sensors at a higher frame rate with respect to other technologies (e.g. CCD devices~\cite{bib:ccd}). 
ORCA-Flash 4.0 camera is provided with a high granularity ($2048\times2048$ pixels), very low noise (standard deviation of readout noise around two photons per pixel and negligible effect of dark current for exposure times lower than 1 second), good sensitivity (70\% quantum efficiency in the 500-700~nm range, comparable with other front-illuminated devices) and good linearity. 

This camera is instrumented with a Schneider lens (25~mm focal length and aperture of f/0.95). The lens is placed at a distance $d$=50.6 cm from the last GEM
in order to obtain a de-magnification
$\delta = (d/f) - 1 = 19.25$ to
image a surface $25.6 \times 25.6$~cm$^2$ onto the
$1.33 \times 1.33$~cm$^2$ sensor. This lens has a 85\%  transmittance on average in the 500 nm to 700 nm light wavelength range.
In this configuration, each pixel
 is therefore imaging  an effective area of 125$\times$125~$\mu$m$^2$ of the GEM layer. Because of the geometrical acceptance~\cite{bib:jinst_orange1}, the fraction of the light collected by the lens can be evaluated to be $1.7 \times 10^{-4}~$. Except from very few pixel rows and columns at the sensor sides, that look outside from the TPC sensitive volume and are removed from the analysis, sensor response and noise level do not show any appreciable dependence on $X$ and $Y$. 

A  cathode has been realized using an ATLAS MicroMegas mesh \cite{bib:mesh}, produced
by Swiss BOOP company with 30~$\mu$m diameter wires with a pitch of 70~$\mu$m, stretched and glued on a ring, 1 cm apart from the first ring of the field cage and ensuring an adequate light transmission. Its transparency to light has been estimated to be about $70\%$.

On this side, LEMOn has been equipped with a 50$\times$50~mm$^2$ HZC Photonics XP3392 photomultiplier \cite{PMTPhotonics} (PMT) detecting light through a transparent $50\times50\times4$~mm$^3$ fused silica window.

This PMT  also allowed to study 
the timing properties of light produced in the innermost GEM layer 250 mm far away during the development of the electron avalanche.  The  field cage is contained in a  $370\times270\times280$~mm$^3$ and 2.5~mm thick box made of ASA. This box, realised to guarantee the gas containment, has been equipped with two 180~$\mu$m thin $200\times200$~mm$^2$ windows made of TEDLAR to reduce as much as possible the multiple scattering of the impinging ultra-relativistic  electrons (or the absorption of particles in other tests).

\section{Detector Operations}

The data reported in this paper have been collected at the BTF which can deliver bunched electrons or positrons from few tens of MeV to several hundred MeV energy. The BTF is optimized to deliver electrons with an energy of 450 MeV with  a typical beam spot size of $\sigma_{x,y}\simeq$~2~mm, a divergence  $\sigma^\prime_{x,y}\simeq$~2~mrad, and an energy spread of $\simeq$~1\%. Moreover, for our measurements the bunch multiplicity has been tuned to be  single particle, with the possibility to change it up to several hundred  electrons (with a 10~ns bunch length). The bunch multiplicity and the beam spot-size were monitored by means of a silicon pixel detector (Fast interface for Timepix pixel detectors, FitPix~\cite{Kraus_2011}) located upstream of LEMOn  and a Pb-glass  calorimeter~\cite{Buonomo:2017sdz, Buonomo:2017btf} accommodated at the back of LEMOn and acquired together with the LEMOn camera image and the PMT signal. All the devices in the acquisition chain were triggered by the DAPHNE accelerator timing system of which the BTF is part \cite{bib:daphne_time}, that allows to distribute the trigger to the camera, 
to the FitPix, and to a LeCroy 610Zi used to digitise at a sampling frequency of 10 GS/s and acquire waveforms from the LEMON-PMT and from the calorimeter.

\begin{figure}[!ht]
\centering
\includegraphics[width=3in]{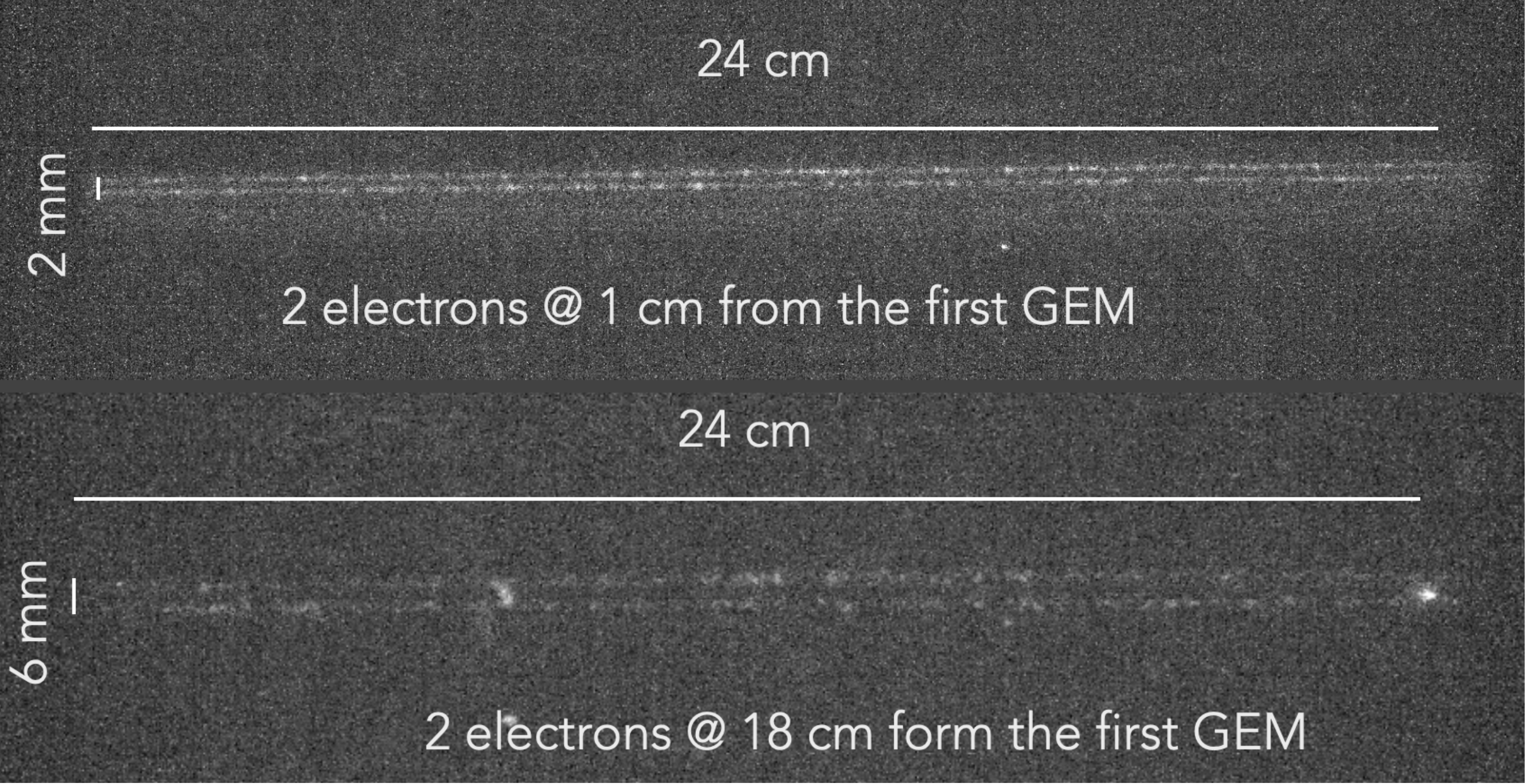}\DeclareGraphicsExtensions.
\caption{Two examples of  two-tracks events collected at the Frascati BTF. The two BTF electrons are crossing the field cage parallel  to  the  240~mm major axis. One event is acquired with the beam close to the GEM (up), and another with the beam 180 mm far away  from the GEM (down). The separation among the two track can be easily measured.}
\label{fig:track}
\end{figure}

LEMOn was operated with a He-CF$_4$ (60/40) gas mixture, with a voltage of 455~V across the sides of each of the three GEMs and an electric field between them of 2.0 kV/cm. The light emission spectrum of CF$_4$ has a substantial component in the 600-700 nm light wavelength range \cite{FRAGA200388}. A gas mixer was used to provide the right mixture starting from pure gases with a precision of 0.5\% for both components. The ratio between the two gases was optimised to provide enough light, while keeping the detector working in stable ans safe electrostatic conditions \cite{bib:stab}.
In this high voltage configuration the GEM stack provides a gain of about $5\times10^5$, considered large enough to reach a good light production in  the avalanche generated in the  GEMs structure. The typical photon yield for this  type of gas mixtures has been measured to be around  0.07 photons per avalanche electron~\cite{bib:jinst_orange1, bib:roby, bib:tesinatalia}.
The gas mixture was kept under a continuous flow of about 200 cc/min (equivalent to a 1.5 gas volume exchange every hour) with a 2 mbar over-pressure over the atmospheric pressure value.

The field cage was powered by a High Voltage Power Supply CAEN N1570\cite{CAENN1570} able to deliver up to 15 kV. 
This posed a limitation to the maximum  electric field  at 0.6~kV/cm. 

The triple GEM system was powered with a HV GEM power supply \cite{Corradi:2007df} ensuring stability and accurate monitoring of the bias currents.

The ORCA Camera I/O was configured to work in {\it Global Reset Mode}. The pixel lines are initialized one at a time. When the sensor is fully initialised, all previously collected charges are reset and all pixels are simultaneously exposed  for the Global Exposure Time, configured by software. 

The initialisation process lasts about 80~$\mu$s and was triggered by a signal properly generated by the accelerator timing system previously described.


Optics and exposure time (30~ms) were optimized to ensure the largest light collection and to reduce as much as possible events due to the natural radioactivity. One hundred images were typically acquired in each detector configuration ({\it run} in the following). 
Fig.~\ref{fig:track} shows two examples of BTF electron tracks images acquired with LEMOn.

The 450 MeV electrons were delivered by the LNF accelerator complex  along the X axis direction (see Fig.\ref{fig:sex})  with a repetition rate of one bunch every  second. LEMOn was accommodated over a remotely controlled table  in order to scan 
the $Z$ coordinate (with a 0.2 mm precision). 


\section{Test Beam Results}

Data were collected during a week long campaign, when LEMOn was continuously operated. 

The fast PMT signal waveform was acquired using as external trigger from the timing signal of the BTF line synchronized with the electrons arrivals. The time $t_{s}$ corresponding to a fixed voltage of the PMT waveform was associated to each PMT signal and data at different longitudinal $Z$ position were collected. 
The standard deviation ($\sigma_t$) of the distribution of the residuals of $t_{s}$ at each $Z$ can be converted into the standard deviation ($\sigma_Z$) of the BTF electrons $Z$ position using the gas mixture drift velocity that was simulated with Garfield \cite{bib:garfield} to be 6~cm/$\mu$s for an electric field of 500~V/cm. A value of $\sigma_Z$ around  1 mm  was found,  well compatible with the  beam spot transverse size. 

Each image acquired with the sCMOS camera was saved as a 2048 x 2048 matrix of photon counts. 
 Because of the very low occupancy of sensor, the baseline noise of the sensor is estimated pixel by pixel by obtaining the distribution of counts for each pixel in all the images of a data-taking run. The  average count and the standard deviation ($\sigma_n$) for each pixel is then evaluated. This average photon count  is subtracted to the count of each image before the image is further processed\cite{bib:fe55}. 
 
 The reconstruction  of tracks in each image is then made by using a Hough transform pattern recognition algorithm (HT).
 The HT is used to detect  straight lines  by converting the representation  $y$ = $m \cdot x + b$ in the ($m$ , $b$ ) parameter space to the $r$ = $x \cos \theta + y \sin \theta$ representation where $r$ is the distance from the origin to the closest point on the straight line, and $ \theta $ is the angle between the $x$ axis and the line connecting the origin with that closest point. Each line in the image can be associated to a pair ($r$, $\theta$) and the problem of finding a straight line is converted in a problem of finding concurrent curves in the ($r$, $\theta$) plane \cite{bib:hough}.
 
 Since LEMOn is positioned to let the BTF electrons cross the drift volume at $Y$ $\sim$ 0, only pixels with a photon count exceeding 1.5$\sigma_n$ are used in the HT. The HT could finds several lines connecting all the pixels above this threshold: the most ranked line (i.e. the one that minimise the sum of residuals to points) within an angle of about $\pm 5$ mrad respect to the $X$ axis is then selected in each image and it represents the candidate reconstructed BTF electron (a {\it track}). Multiple track images are also analyzed and the HT is in fact able to distinguish  tracks in events with multiplicity larger than one.

The efficiency to reconstruct an electron crossing the 24 cm wide drift region  was estimated by selecting a sample of events with the calorimeter signal compatible with a single BTF  electron  deposit. In all the events considered  the HT was always able to find a track in the sCMOS image. This is well compatible with a Garfield \cite{bib:garfield1, bib:garfield} simulation of the  gas mixture that yields an  estimate for the average  energy loss per single ionization of 42~eV and about 5 primary electrons produced per millimeter, resulting in a dE/dx of about 0.21~keV/mm for a $\simeq$~450 MeV electron. This energy loss translates into three primary ionization e$^-$ cluster per track mm. The sequence of several ionization clusters along the electron trajectory is  a  signature  clear enough to detect each electron with the HT.

\subsection{Position and energy resolutions}

\begin{figure}[!ht]
\centering
\includegraphics[width=3.4in]{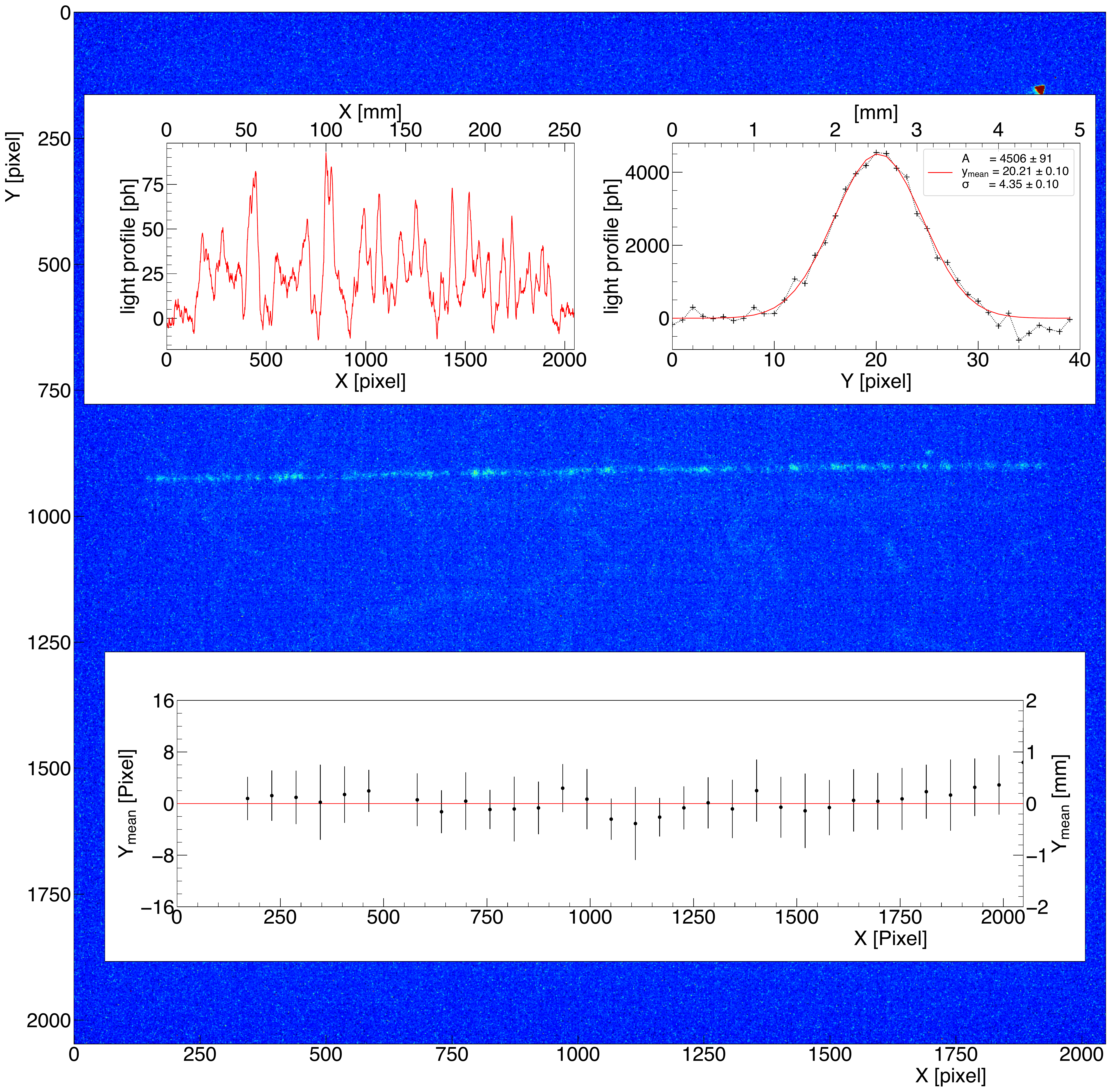}\DeclareGraphicsExtensions.
\caption{{\bf Background}: sCMOS camera  image collected at the BTF with beam passing 6.5 cm apart from GEM plane. Pixels collecting light due to a BTF electron are visible at $Y$ $\sim$ 1000. {\bf Left top insert}:  light distribution (after pixel-by-pixel noise subtraction) along the track $X$-direction summing all the photons for 40 pixel in the  $Y$ direction around the track direction. {\bf Right top insert}: light distribution  transverse to the track ($Y$ direction) summing all the photons along the  $X$ direction  with a superimposed Gaussian fit. {\bf Bottom insert}: Gaussian $Y_{mean}$ of the transverse light distribution 
with superimposed the line found by the HT (subtracted of the $Y$ of the line). 
The error bar of the points is the sigma of the normal fit in each slice and it is taken as a indication of the width of the light deposit. Some segments with a  too low signal-to-noise ratio of the detected light are not displayed. }
\label{fig:tracking}
\end{figure}

 The profile of light distribution transverse to the track direction is evaluated for each track and fitted to a Gaussian function (right top insert of Fig. \ref{fig:tracking}).
 Its integral represents the total light yield for the track and is therefore proportional its total energy loss in the gas volume. 

 In all analyses presented in this paper, tracks have been divided in 36 portions about 7 mm long. 
 
 We evaluate the performance of LEMOn in measuring their characteristics: position in space, light content, longitudinal and transverse light profile. 
 An average energy loss of about 1.5 keV corresponds to each segment (with a production of 21 clusters for a total of almost 40 primary electrons per segment) allowing to study the performance of LEMOn to in the keV energy release range.
 
   Instrumental effects such as electric field distortion (due to the coupling between an elliptical field cage and rectangular GEMs) and optical vignetting, have been observed in the external regions of sensitive volume of LEMOn.
  Different methods to perform quantitative analyses of these effects are now under study. They are expected to be completely deterministic and, once studied, they could be off-line corrected and should not represent a major issue on larger prototypes.
  For the results presented in this paper, among the 36 segments, we decided to retain only the 18 in the central region of the field cage  ellipse, where detector response was found to be uniform, for subsequent analysis.

\begin{figure}[!ht]
\centering
\includegraphics[width=3.4in]{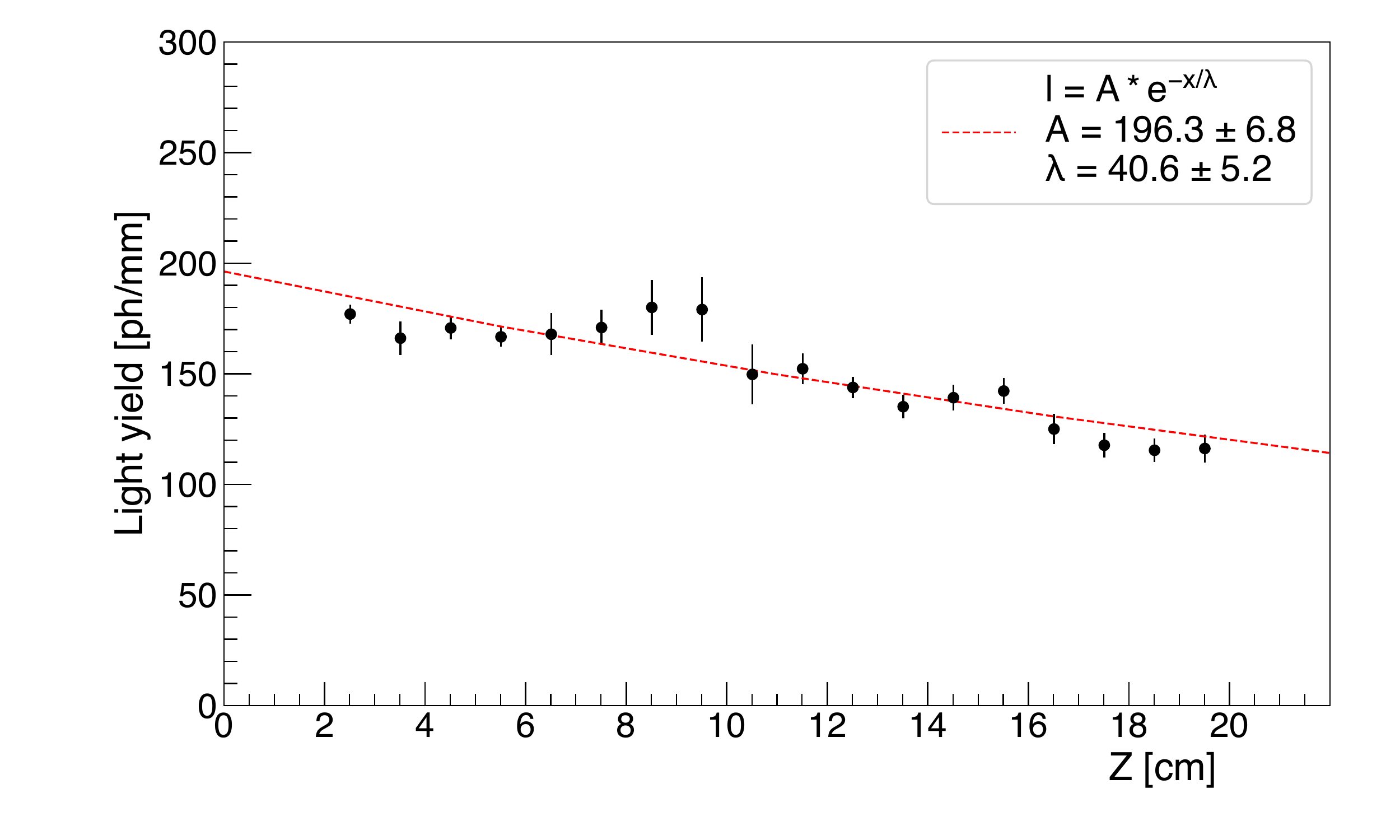}
\caption{Average light yield $I$ per track as a function of the track $Z$ position across the field cage.}
\label{fig:lightvsZ}
\end{figure}

In different runs the LEMOn position along $Z$ is changed resulting in a different $Z$ coordinate for the tracks. These runs  are used to evaluate the uniformity versus the drift axis  (Z) of the TPC. The light yield shows a decrease at larger $Z$ positions (see Fig.\ref{fig:lightvsZ}) resulting in
an absorption length $\lambda$ of about~40~cm, as evaluated from the fit. 
 A very similar decrease of the light yield was also measured by the PMT, excluding this to be due to some instrumental or analysis effect.
 This is likely due to the electron attachment to gas impurities during the drift of the ionization electrons from their production points along the track to the GEMs. We did not have a direct measurement of gas contaminants. If the light reduction were due to oxygen, the measured value of $\lambda$ would indicate a contamination of about 300 ppm \cite{bib:2001mug}.

\begin{figure}[!ht]
\centering
\includegraphics[width=3.4in]{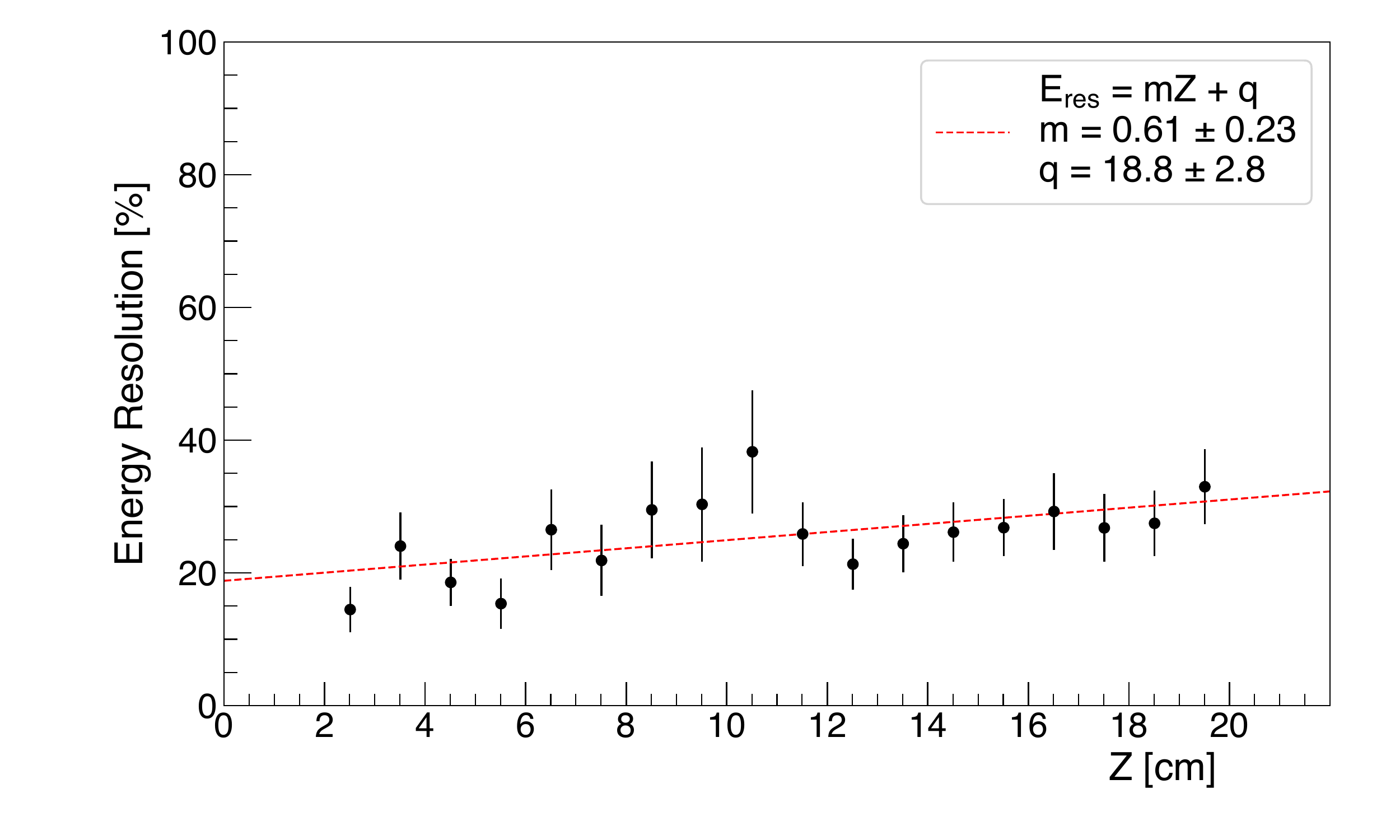}
\caption{Energy Resolution: standard deviation of the light yield divided by its average as a function of the track $Z$ position.}
\label{fig:energyresvsZ}
\end{figure}

  Moreover, the relative fluctuation of the light yield value represents an estimate of the energy release resolution (Fig.\ref{fig:energyresvsZ}). This turns out to have a slight dependence on the $Z$ position of the track increasing from about 20\% at $Z$=0~cm to about 30\% at $Z$=20~cm, corresponding to an absolute resolution between 300~eV and 450~eV.
  Since an average number of about 21 electron clusters per segment are produced, this result can almost completely be explained with the statistical fluctuations of gas ionization.

\begin{figure}[!ht]
\centering
\includegraphics[width=3.4in]{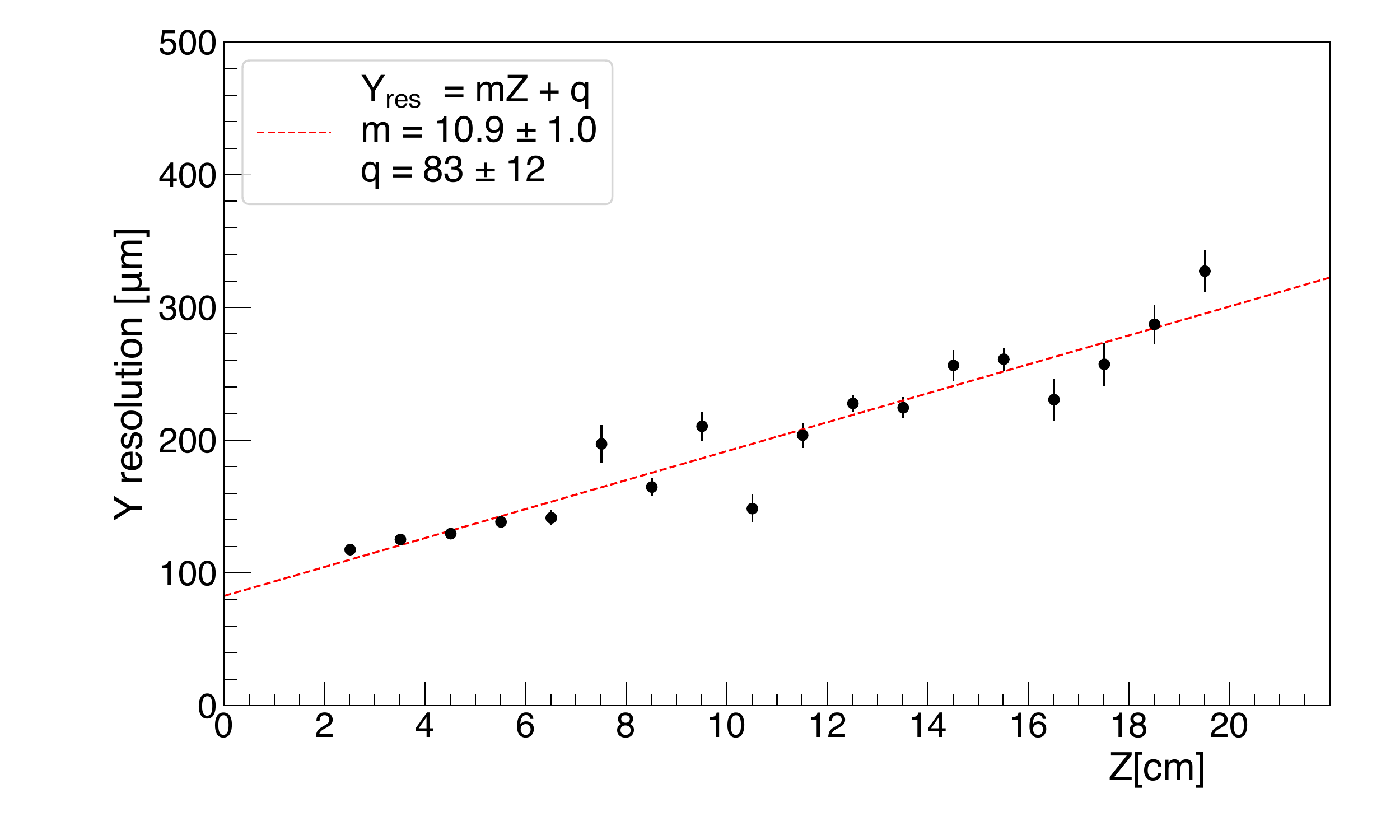}\DeclareGraphicsExtensions.
\caption{ Segment $Y$ position resolution  as a function of the track's $Z$ coordinate.}
\label{fig:XYres}
\end{figure}

Also the transverse  light profile of each segment can be described by  a Gaussian distribution. The fitted mean value  $Y_{mean}$ of this distribution  is used to identify the segment's $Y$ position. In Fig. \ref{fig:tracking}   examples of longitudinal and transverse light profiles are shown. 
  The distribution of the residuals of $Y_{mean}$ to the line obtained with HT is then considered for the 18 segments. The standard deviation of this distribution evaluated in a sample of several tracks represents an estimate of the resolution on the segment's $Y$ position. This procedure is repeated for several images acquired with the BTF electrons  crossing the LEMOn field cage at different $Z$ positions (Fig.\ref{fig:XYres}).
The  dependence of the segment's $Y$ position resolution  on the $Z$ track coordinate is then interpolated with a linear function (Fig.\ref{fig:XYres}). A resolution  of 83 $\pm$ 12 $\mu$m is extrapolated close to the GEMs ($Z$ = 0), in agreement with an almost three times better result obtained with an almost three times smaller effective pixel in a previous work~\cite{bib:ieee_orange}.
We observe its worsening for larger $Z$: this is mainly due to the reduction of the number of electrons, probably due to re-absorption, for larger $Z$ that, besides reducing the produced light, also amplifies the effect of diffusion.

\subsection{$Z$ coordinate measurement}

\begin{figure}[!ht]
\centering
\includegraphics[width=3.4in]{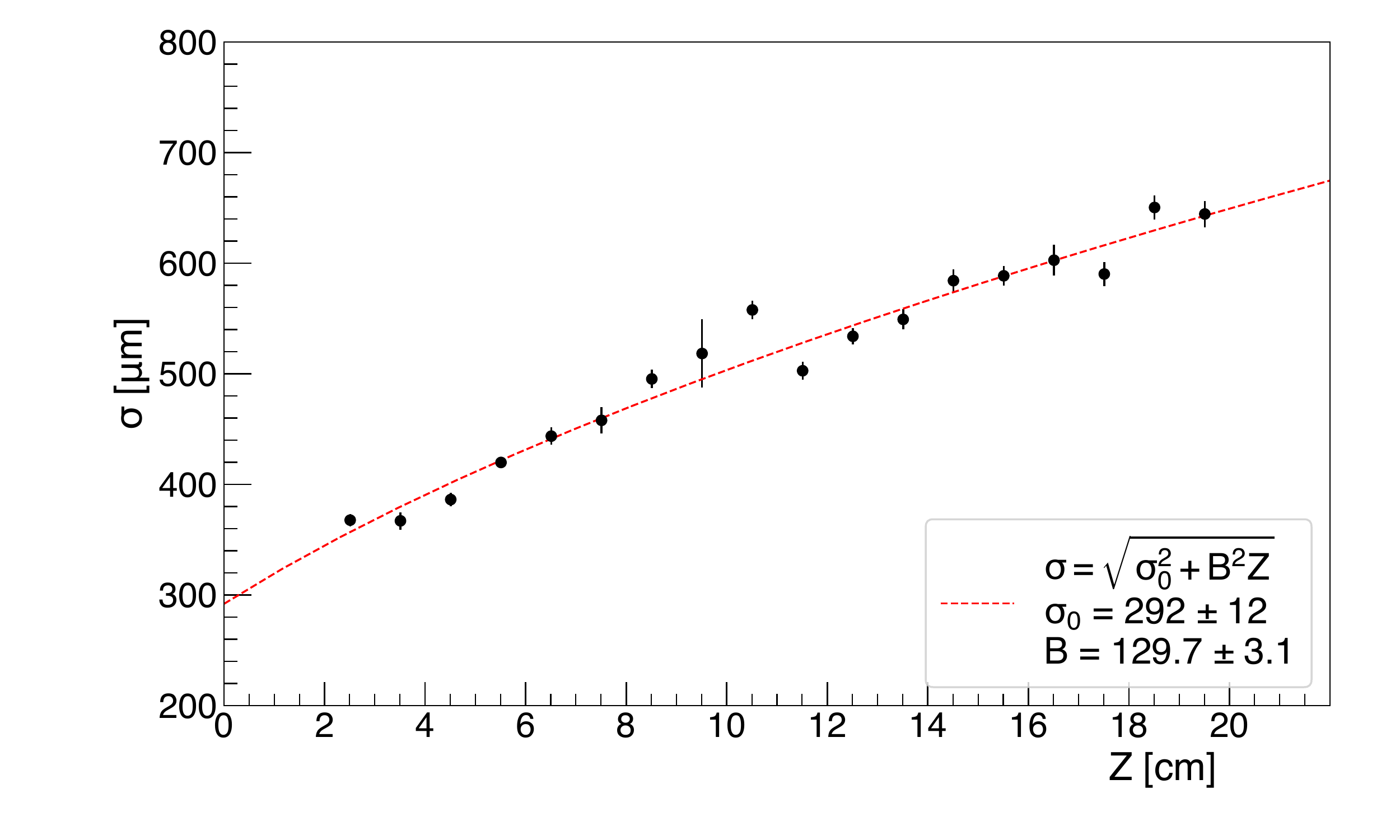}\DeclareGraphicsExtensions.
\caption{ Average $\sigma$ of transverse light distribution for track segments as a function of the track $Z $ coordinate. }
\label{fig:Diffusion}
\end{figure}

 During the drift in the gas, electrons are subject to longitudinal and transverse diffusion effects, 
 that influence their arrival $X$ and $Y$ coordinates and their arrival time at the GEMs.
 For each single electron, the position at the anode has a probability distribution function that can be described by a 2D Gaussian with $\sigma_X$  and $\sigma_Y$ equal to $B~=~\sqrt{\frac{2DZ}{\mu E}}$ where $D$ is the diffusion coefficient, $\mu$ the electron mobility and $E$ the drift electric field \cite{bib:rolandiblum}. Moreover, during the avalanche formation within the GEMs a further diffusion of the avalanche electrons is taking place. Eventually, the light recorded by the sCMOS camera and by the PMT is related to the original point with an uncertainty that is larger for production points farther in $Z$ from the GEMs.
  This is reflected in the transverse ($Y$) light distribution of each segment of the track: the $\sigma_Y$  obtained from  the  Gaussian fit is in fact increasing with $\sqrt{Z}$. Therefore, the segment's original $Z$ can be deduced by measuring  $\sigma$(Fig.\ref{fig:Diffusion}).
  
 The observed values of $\sigma$  are related to the track's $Z$  by  $\sigma = \sqrt{\sigma_0^2 + B^2 Z }$ 
 
 The transverse  diffusion coefficient $B$ in the gas is measured to be $129.7 \pm 3.1$ $\frac{\mu m}{\sqrt{cm}}$,  well in agreement with the expected value obtained with Garfield simulation (130 $\frac{\mu m}{\sqrt{cm}}$)~\cite{bib:stab}.

 The intercept at zero $\sigma_0$ =  $292 \pm 12$ $\mu$m is due to the contribution of the electron avalanche propagation in the GEM stack and confirms results obtained with a radioactive source~\cite{bib:stab}.
 
From the same Gaussian fit to the light  $Y$ distribution of each segment, the amplitude $A$ can be obtained. Since $\sigma A$ is proportional to the total light $I$ of the segment, we can  define $\eta = \frac{\sigma}{A}$ that is therefore  proportional to $\frac{\sigma_0^2 + B^2 Z}{I}$.

\begin{figure}[ht]
\centering
\includegraphics[width=3.1in]{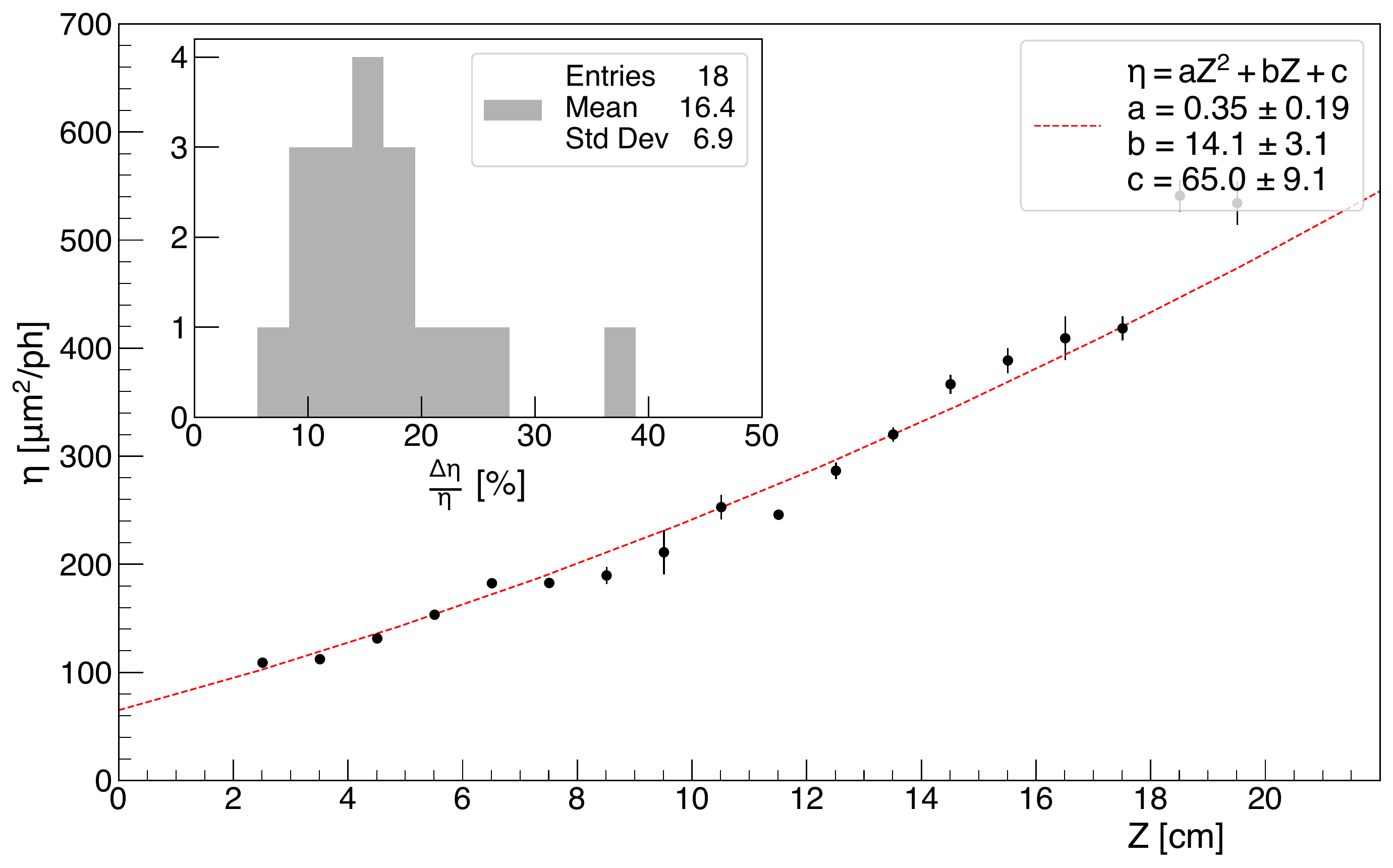}\DeclareGraphicsExtensions.
\caption{ Average $\eta$ for track segments as a function of the track's $Z$ coordinate with a quadratic fit superimposed. Relative uncertainty on $\eta$ (inset) for the various $Z$ positions. }
\label{fig:etavsZ}
\end{figure}

In Fig.~\ref{fig:etavsZ} we show the dependence of $\eta$ 
on the track's $Z$ coordinate. Since in our data   the light $I$ shows an exponential decrease with $Z$ (see Fig.\ref{fig:lightvsZ}), $\eta$ behavior can be approximated as:

\begin{equation}
\eta~\simeq~\frac{(\sigma_0^2 + B^2 Z)}{I_0} \cdot (1+ Z/\lambda)
\end{equation}

Therefore, a quadratic fit is expected to give a good representation of the data.
 
\begin{figure}[ht]
\centering
\includegraphics[width=3.4in]{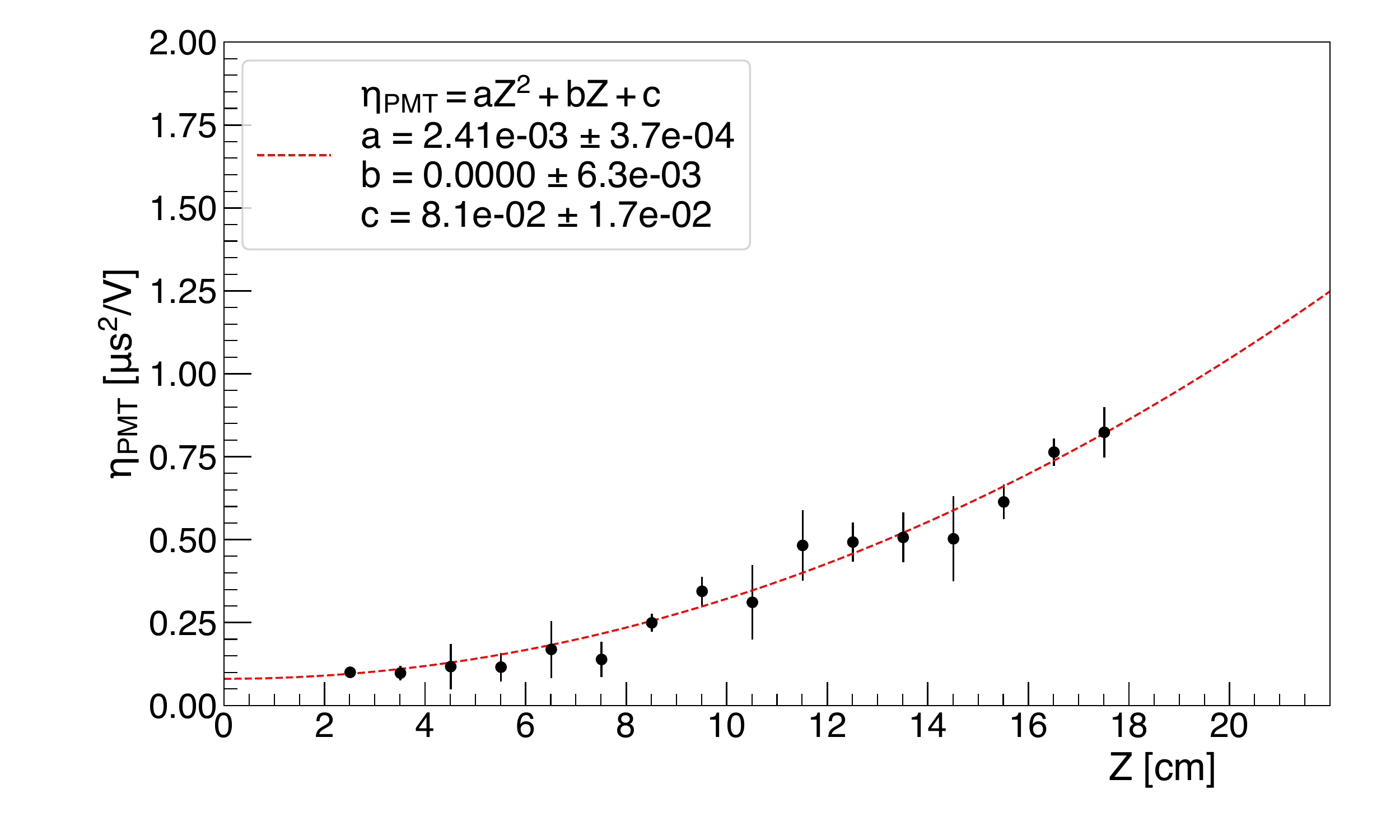}\DeclareGraphicsExtensions.
\caption{Average $\eta_{PMT}$ as a function of the track $Z$ coordinate. Data were not recorded for the largest $Z$ point since the beam was adding extra noise in the PMT.}
\label{fig:Tdiffusion}
\end{figure}

A similar parameter $\eta_{PMT}$ can be defined from the analysis of the PMT waveform. In this case the total light of the track is recorded (including the light from the segments excluded from the sCMOS camera data analysis).  The Pb-glass calorimeter signal of the BTF line was however used to reject  events with  more than one  track. 
The amplitude of the  PMT waveform and its width can be similarly  used to calculate $\eta_{PMT}$. In this case the width of the waveform is larger for more distant tracks (larger $Z$)  due to the longitudinal diffusion of the drifting ionization electrons. By using the drift velocity, $\eta_{PMT}$ can be related to $Z$ similarly to $\eta$ (Fig.\ref{fig:Tdiffusion}). 

Transverse  and longitudinal  diffusion can therefore  be exploited to measure the longitudinal $Z$ coordinates. We can estimate $\frac{\Delta \eta}{\eta}$  from the standard deviation of the distributions of the  $\eta$ values. Since $\frac{\Delta \eta}{\eta} =\frac{\Delta Z}{Z} $, the latter can be used to estimate  $\frac{\Delta Z}{Z}$. For each single segment, this resolution turns out to be  in the range 10\% - 20\% for both methods, even if measurements with sCMOS lead to a slightly better evaluation of $Z$.

Moreover, a combination of the $\eta$ parameters (from sCMOS and PMT) would probably give an even better evaluation $Z$ that would result to be very  useful to define a fiducial region  of a DM detector. The anode where the GEMs are located (at $Z$ $\sim$ 0 in LEMOn) and the cathode (at $Z$ $\sim$ 20 cm in LEMOn) are usually the most radioactive elements in a DM detector. They are in fact sources of spurious nuclear recoils that would be easily  removed if their $Z$ coordinate is measured.

\section{Conclusion}

The data collected at the Frascati BTF with the LEMOn prototype, part of the CYGNO project, are confirming the potentiality of large optically readout TPC as detector for low energy events.

In this paper we have reported how  ultra-relativistic electron tracks can be  very efficiently reconstructed by collecting the light emitted from GEMs with a high resolution and high sensitivity sCMOS camera and a PMT.
Particle trajectories were found by means of an Hough transform method that worked very well on long and straight tracks. For curly and short ones a different method was developed and successfully applied to the analysis of images with low energy particle interactions \cite{bib:clus}.

The analysis of the 7 mm long slices of an electron track shows a  good spatial  and  energy resolution,  making very promising the use of this  gas TPC down to few keV energy releases. Also a method based on the ionization electron diffusion confirmed to be very effective in determining keV segment's longitudinal position within the field cage. The combination of the PMT and camera readout can offer a 3D reconstruction of the particles tracks as demonstrated in smaller size prototypes (\cite{bib:jinst_orange2}).

The very good sensitivities measured in the low energy region make this technology extremely interesting in the development of a larger scale detector operated at atmospheric pressure aiming to observe very rare processes as Dark Matter or Solar Neutrinos interactions with ordinary matter.

\section*{Acknowledgment}
A special thanks goes to M. Iannarelli, now retired, who assembled LEMOn and we are grateful to the BTF staff for the excellent and stable beam condition and their inexhaustible support.

This work was partially supported by the European Research Council (ERC) under the European Union’s Horizon 2020 research and innovation program (grant agreement No 818744).

\section*{Data Availability Statement}
Data available on request from the authors.

\nocite{*}
\bibliography{main}

\providecommand{\noopsort}[1]{}\providecommand{\singleletter}[1]{#1}%
\begin{thebibliography}{10}
\expandafter\ifx\csname url\endcsname\relax
  \def\url#1{\texttt{#1}}\fi
\expandafter\ifx\csname urlprefix\endcsname\relax\def\urlprefix{URL }\fi
\expandafter\ifx\csname href\endcsname\relax
  \def\href#1#2{#2} \def\path#1{#1}\fi

\bibitem{bib:tpc1}
J.~Alme, et~al., {The ALICE TPC, a large 3-dimensional tracking device with
  fast readout for ultra-high multiplicity events}, Nucl. Instrum. Meth. A 622
  (2010) 316--367.
\newblock \href {http://arxiv.org/abs/1001.1950} {\path{arXiv:1001.1950}},
  \href {https://doi.org/10.1016/j.nima.2010.04.042}
  {\path{doi:10.1016/j.nima.2010.04.042}}.

\bibitem{bib:tpc2}
C.~Rubbia, et~al., {Underground operation of the ICARUS T600 LAr-TPC: first
  results}, JINST 6 (2011) P07011.
\newblock \href {http://arxiv.org/abs/1106.0975} {\path{arXiv:1106.0975}},
  \href {https://doi.org/10.1088/1748-0221/6/07/P07011}
  {\path{doi:10.1088/1748-0221/6/07/P07011}}.

\bibitem{bib:tpc3}
K.~Terao, {MicroBooNE: Liquid Argon TPC at Fermilab}, JPS Conf. Proc. 8 (2015)
  023014.
\newblock \href {https://doi.org/10.7566/JPSCP.8.023014}
  {\path{doi:10.7566/JPSCP.8.023014}}.

\bibitem{bib:tpc4}
O.~Barring, {Studies of the ionisation energy loss in the DELPHI TPC and the
  identification of quark and gluon jets in hadronic events at LEP}, Other
  thesis (10 1992).

\bibitem{Battat:2016xxe}
J.~Battat, et~al., {Low Threshold Results and Limits from the DRIFT Directional
  Dark Matter Detector}, Astropart.\ Phys. 91 (2017) 65--74.
\newblock \href {http://arxiv.org/abs/1701.00171} {\path{arXiv:1701.00171}},
  \href {https://doi.org/10.1016/j.astropartphys.2017.03.007}
  {\path{doi:10.1016/j.astropartphys.2017.03.007}}.

\bibitem{BATTAT20151}
J.~Battat, J.~Brack, E.~Daw, A.~Dorofeev, A.~Ezeribe, J.-L. Gauvreau, M.~Gold,
  J.~Harton, J.~Landers, E.~Law, E.~Lee, D.~Loomba, A.~Lumnah, J.~Matthews,
  E.~Miller, A.~Monte, F.~Mouton, A.~Murphy, S.~Paling, N.~Phan, M.~Robinson,
  S.~Sadler, A.~Scarff, F.~{Schuckman II}, D.~Snowden-Ifft, N.~Spooner,
  S.~Telfer, S.~Vahsen, D.~Walker, D.~Warner, L.~Yuriev,
  \href{http://www.sciencedirect.com/science/article/pii/S2212686415000084}{First
  background-free limit from a directional dark matter experiment: Results from
  a fully fiducialised drift detector}, Physics of the Dark Universe 9-10
  (2015) 1 -- 7.
\newblock \href {https://doi.org/https://doi.org/10.1016/j.dark.2015.06.001}
  {\path{doi:https://doi.org/10.1016/j.dark.2015.06.001}}.
\newline\urlprefix\url{http://www.sciencedirect.com/science/article/pii/S2212686415000084}

\bibitem{Ikeda:2020mvr}
T.~Ikeda, K.~Miuchi, T.~Hashimoto, H.~Ishiura, T.~Nakamura, T.~Shimada,
  K.~Nakamura, {Results of a directional dark matter search from the NEWAGE
  experiment}, J. Phys. Conf. Ser. 1468~(1) (2020) 012042.
\newblock \href {https://doi.org/10.1088/1742-6596/1468/1/012042}
  {\path{doi:10.1088/1742-6596/1468/1/012042}}.

\bibitem{Battat:2016pap}
J.~Battat, et~al., {Readout technologies for directional WIMP Dark Matter
  detection}, Phys.\ Rept. 662 (2016) 1--46.
\newblock \href {http://arxiv.org/abs/1610.02396} {\path{arXiv:1610.02396}},
  \href {https://doi.org/10.1016/j.physrep.2016.10.001}
  {\path{doi:10.1016/j.physrep.2016.10.001}}.

\bibitem{ALNER2005173}
G.~Alner, H.~Araujo, A.~Bewick, S.~Burgos, M.~Carson, J.~Davies, E.~Daw,
  J.~Dawson, J.~Forbes, T.~Gamble, M.~Garcia, C.~Ghag, M.~Gold, S.~Hollen,
  R.~Hollingworth, A.~Howard, J.~Kirkpatrick, V.~Kudryavtsev, T.~Lawson,
  V.~Lebedenko, J.~Lewin, P.~Lightfoot, I.~Liubarsky, D.~Loomba, R.~Lüscher,
  J.~McMillan, B.~Morgan, D.~Muna, A.~Murphy, G.~Nicklin, S.~Paling, A.~Petkov,
  S.~Plank, R.~Preece, J.~Quenby, M.~Robinson, N.~Sanghi, N.~Smith, P.~Smith,
  D.~Snowden-Ifft, N.~Spooner, T.~Sumner, D.~Tovey, J.~Turk, E.~Tziaferi,
  R.~Walker,
  \href{http://www.sciencedirect.com/science/article/pii/S0168900205018139}{The
  drift-ii dark matter detector: Design and commissioning}, Nuclear Instruments
  and Methods in Physics Research Section A: Accelerators, Spectrometers,
  Detectors and Associated Equipment 555~(1) (2005) 173 -- 183.
\newblock \href {https://doi.org/https://doi.org/10.1016/j.nima.2005.09.011}
  {\path{doi:https://doi.org/10.1016/j.nima.2005.09.011}}.
\newline\urlprefix\url{http://www.sciencedirect.com/science/article/pii/S0168900205018139}

\bibitem{Riffard:2016mgw}
Q.~Riffard, et~al., {MIMAC low energy electron-recoil discrimination measured
  with fast neutrons}, JINST 11~(08) (2016) P08011.
\newblock \href {http://arxiv.org/abs/1602.01738} {\path{arXiv:1602.01738}},
  \href {https://doi.org/10.1088/1748-0221/11/08/P08011}
  {\path{doi:10.1088/1748-0221/11/08/P08011}}.

\bibitem{Sauzet:2020dut}
N.~Sauzet, D.~Santos, O.~Guillaudin, G.~Bosson, J.~Bouvier, T.~Descombes,
  M.~Marton, J.~Muraz, {Fast neutron spectroscopy with Mimac-FastN : a mobile
  and directional fast neutron spectrometer, from 1 MeV up to 15 MeV}, J. Phys.
  Conf. Ser. 1498~(1) (2020) 012044.
\newblock \href {https://doi.org/10.1088/1742-6596/1498/1/012044}
  {\path{doi:10.1088/1742-6596/1498/1/012044}}.

\bibitem{Hashimoto:2017hlz}
T.~Hashimoto, K.~Miuchi, K.~Nakamura, R.~Yakabe, T.~Ikeda, R.~Taishaku,
  M.~Nakazawa, H.~Ishiura, A.~Ochi, Y.~Takeuchi, {Development of a
  low-alpha-emitting $\mu$-PIC for NEWAGE direction-sensitive dark-matter
  search}, AIP Conf. Proc. 1921~(1) (2018) 070001.
\newblock \href {http://arxiv.org/abs/1707.09744} {\path{arXiv:1707.09744}},
  \href {https://doi.org/10.1063/1.5019004} {\path{doi:10.1063/1.5019004}}.

\bibitem{bib:vahsen}
S.~Vahsen, K.~Oliver-Mallory, M.~Lopez-Thibodeaux, J.~Kadyk,
  M.~Garcia-Sciveres, {Tests of gases in a mini-TPC with pixel chip readout},
  Nucl. Instrum. Meth. A 738 (2014) 111--118.
\newblock \href {https://doi.org/10.1016/j.nima.2013.10.029}
  {\path{doi:10.1016/j.nima.2013.10.029}}.

\bibitem{Seguinot:1992zu}
J.~Seguinot, T.~Ypsilantis, A.~Zichichi, {A High rate solar neutrino detector
  with energy determination}, Conf.\ Proc.\ C 920310 (1992) 289--313.

\bibitem{ARPESELLA1996333}
C.~Arpesella, C.~Broggini, C.~Cattadori,
  \href{http://www.sciencedirect.com/science/article/pii/0927650595000518}{A
  possible gas for solar neutrino spectroscopy}, Astroparticle Physics 4~(4)
  (1996) 333 -- 341.
\newblock \href {https://doi.org/https://doi.org/10.1016/0927-6505(95)00051-8}
  {\path{doi:https://doi.org/10.1016/0927-6505(95)00051-8}}.
\newline\urlprefix\url{http://www.sciencedirect.com/science/article/pii/0927650595000518}

\bibitem{bib:Fraga}
M.~M. F.~R. Fraga, F.~A.~F. Fraga, S.~T.~G. Fetal, L.~M.~S. Margato,
  R.~Ferreira-Marques, A.~J. P.~L. Policarpo, {The GEM scintillation in He
  CF$_4$, Ar CF$_4$, Ar TEA and Xe TEA mixtures}, Nucl. Instrum. Meth. A504
  (2003) 88--92.
\newblock \href {https://doi.org/10.1016/S0168-9002(03)00758-7}
  {\path{doi:10.1016/S0168-9002(03)00758-7}}.

\bibitem{bib:Margato1}
A.~Morozov, L.~M.~S. Margato, M.~M. F.~R. Fraga, L.~Pereira, F.~A.~F. Fraga,
  {Secondary scintillation in CF$_4$: emission spectra and photon yields for
  MSGC and GEM}, JINST 7 (2012) P02008.
\newblock \href {https://doi.org/10.1088/1748-0221/7/02/P02008}
  {\path{doi:10.1088/1748-0221/7/02/P02008}}.

\bibitem{bib:Margato2}
L.~M.~S. Margato, A.~Morozov, M.~M. F.~R. Fraga, L.~Pereira, F.~A.~F. Fraga,
  {Effective decay time of CF$_4$ secondary scintillation}, JINST 8 (2013)
  P07008.
\newblock \href {https://doi.org/10.1088/1748-0221/8/07/P07008}
  {\path{doi:10.1088/1748-0221/8/07/P07008}}.

\bibitem{Buckland}
K.~N. Buckland, M.~J. Lehner, G.~E. Masek, M.~Mojaver,
  \href{https://link.aps.org/doi/10.1103/PhysRevLett.73.1067}{Low pressure
  gaseous detector for particle dark matter}, Phys. Rev. Lett. 73 (1994)
  1067--1070.
\newblock \href {https://doi.org/10.1103/PhysRevLett.73.1067}
  {\path{doi:10.1103/PhysRevLett.73.1067}}.
\newline\urlprefix\url{https://link.aps.org/doi/10.1103/PhysRevLett.73.1067}

\bibitem{Deaconu:2017vam}
C.~Deaconu, M.~Leyton, R.~Corliss, G.~Druitt, R.~Eggleston, N.~Guerrero,
  S.~Henderson, J.~Lopez, J.~Monroe, P.~Fisher, {Measurement of the directional
  sensitivity of Dark Matter Time Projection Chamber detectors}, Phys. Rev. D
  95~(12) (2017) 122002.
\newblock \href {http://arxiv.org/abs/1705.05965} {\path{arXiv:1705.05965}},
  \href {https://doi.org/10.1103/PhysRevD.95.122002}
  {\path{doi:10.1103/PhysRevD.95.122002}}.

\bibitem{BATTAT20146}
J.~B. Battat, C.~Deaconu, G.~Druitt, R.~Eggleston, P.~Fisher, P.~Giampa,
  V.~Gregoric, S.~Henderson, I.~Jaegle, J.~Lawhorn, J.~P. Lopez, J.~Monroe,
  K.~A. Recine, A.~Strandberg, H.~Tomita, S.~Vahsen, H.~Wellenstein,
  \href{http://www.sciencedirect.com/science/article/pii/S016890021400388X}{The
  dark matter time projection chamber 4shooter directional dark matter
  detector: Calibration in a surface laboratory}, Nuclear Instruments and
  Methods in Physics Research Section A: Accelerators, Spectrometers, Detectors
  and Associated Equipment 755 (2014) 6 -- 19.
\newblock \href {https://doi.org/https://doi.org/10.1016/j.nima.2014.04.010}
  {\path{doi:https://doi.org/10.1016/j.nima.2014.04.010}}.
\newline\urlprefix\url{http://www.sciencedirect.com/science/article/pii/S016890021400388X}

\bibitem{bib:pmt1}
P.~Fonte, A.~Breskin, G.~Charpak, W.~Dominik, F.~Sauli, {Beam Test of an
  Imaging High Density Projection Chamber}, Nucl. Instrum. Meth. A 283 (1989)
  658--664.
\newblock \href {https://doi.org/10.1016/0168-9002(89)91436-8}
  {\path{doi:10.1016/0168-9002(89)91436-8}}.

\bibitem{Sauli:1997qp}
F.~Sauli, {GEM: A new concept for electron amplification in gas detectors},
  Nucl.\ Instrum.\ Meth.\ A 386 (1997) 531--534.
\newblock \href {https://doi.org/10.1016/S0168-9002(96)01172-2}
  {\path{doi:10.1016/S0168-9002(96)01172-2}}.

\bibitem{bib:loomba55Fe}
N.~Phan, E.~Lee, D.~Loomba, {Imaging $^{55}$Fe Electron Tracks in a GEM-based
  TPC Using a CCD Readout}, JINST 15~(05) (2020) P05012.
\newblock \href {http://arxiv.org/abs/1703.09883} {\path{arXiv:1703.09883}},
  \href {https://doi.org/10.1088/1748-0221/15/05/P05012}
  {\path{doi:10.1088/1748-0221/15/05/P05012}}.

\bibitem{bib:opto1}
M.~{Cwiok}, W.~{Dominik}, Z.~{Janas}, A.~{Korgul}, K.~{Miernik}, M.~{Pfutzner},
  M.~{Sawicka}, A.~{Wasilewski}, Optical time projection chamber for imaging of
  two-proton decay of /sup 45/fe nucleus, IEEE Transactions on Nuclear Science
  52~(6) (2005) 2895--2899.

\bibitem{bib:opto2}
M.~Pomorski, M.~Pf\"utzner, W.~Dominik, R.~Grzywacz, A.~Stolz, T.~Baumann,
  J.~S. Berryman, H.~Czyrkowski, R.~D\k{a}browski, A.~Fija\l{}kowska,
  T.~Ginter, J.~Johnson, G.~Kami\ifmmode~\acute{n}\else \'{n}\fi{}ski,
  N.~Larson, S.~N. Liddick, M.~Madurga, C.~Mazzocchi, S.~Mianowski, K.~Miernik,
  D.~Miller, S.~Paulauskas, J.~Pereira, K.~P. Rykaczewski, S.~Suchyta,
  \href{https://link.aps.org/doi/10.1103/PhysRevC.90.014311}{Proton
  spectroscopy of $^{48}\mathrm{Ni},^{46}\mathrm{Fe}$, and $^{44}\mathrm{Cr}$},
  Phys. Rev. C 90 (2014) 014311.
\newblock \href {https://doi.org/10.1103/PhysRevC.90.014311}
  {\path{doi:10.1103/PhysRevC.90.014311}}.
\newline\urlprefix\url{https://link.aps.org/doi/10.1103/PhysRevC.90.014311}

\bibitem{BRUNBAUER201824}
F.~Brunbauer, G.~Galgóczi, D.~{Gonzalez Diaz}, E.~Oliveri, F.~Resnati,
  L.~Ropelewski, C.~Streli, P.~Thuiner, M.~{van Stenis},
  \href{http://www.sciencedirect.com/science/article/pii/S016890021731495X}{Live
  event reconstruction in an optically read out gem-based tpc}, Nuclear
  Instruments and Methods in Physics Research Section A: Accelerators,
  Spectrometers, Detectors and Associated Equipment 886 (2018) 24 -- 29.
\newblock \href {https://doi.org/https://doi.org/10.1016/j.nima.2017.12.077}
  {\path{doi:https://doi.org/10.1016/j.nima.2017.12.077}}.
\newline\urlprefix\url{http://www.sciencedirect.com/science/article/pii/S016890021731495X}

\bibitem{bib:jinst_orange1}
M.~Marafini, V.~Patera, D.~Pinci, A.~Sarti, A.~Sciubba, E.~Spiriti, {High
  granularity tracker based on a Triple-GEM optically read by a CMOS-based
  camera}, JINST 10~(12) (2015) P12010.
\newblock \href {http://arxiv.org/abs/1508.07143} {\path{arXiv:1508.07143}},
  \href {https://doi.org/10.1088/1748-0221/10/12/P12010}
  {\path{doi:10.1088/1748-0221/10/12/P12010}}.

\bibitem{bib:nim_orange2}
M.~Marafini, V.~Patera, D.~Pinci, A.~Sarti, A.~Sciubba, E.~Spiriti,
  \href{http://www.sciencedirect.com/science/article/pii/S0168900215014230}{Optical
  readout of a triple-gem detector by means of a cmos sensor}, Nuclear
  Instruments and Methods in Physics Research Section A: Accelerators,
  Spectrometers, Detectors and Associated Equipment 824 (2016) 562 -- 564,
  frontier Detectors for Frontier Physics: Proceedings of the 13th Pisa Meeting
  on Advanced Detectors.
\newblock \href {https://doi.org/https://doi.org/10.1016/j.nima.2015.11.058}
  {\path{doi:https://doi.org/10.1016/j.nima.2015.11.058}}.
\newline\urlprefix\url{http://www.sciencedirect.com/science/article/pii/S0168900215014230}

\bibitem{JINST:nitec}
E.~Baracchini, G.~Cavoto, G.~Mazzitelli, F.~Murtas, F.~Renga, S.~Tomassini,
  \href{https://doi.org/10.1088%2F1748-0221%2F13%2F04%2Fp04022}{Negative ion
  time projection chamber operation with {SF}6at nearly atmospheric pressure},
  Journal of Instrumentation 13~(04) (2018) P04022--P04022.
\newblock \href {https://doi.org/10.1088/1748-0221/13/04/p04022}
  {\path{doi:10.1088/1748-0221/13/04/p04022}}.
\newline\urlprefix\url{https://doi.org/10.1088%2F1748-0221%2F13%2F04%2Fp04022}

\bibitem{NIM:Marafinietal}
M.~Marafini, V.~Patera, D.~Pinci, A.~Sarti, A.~Sciubba, E.~Spiriti, {ORANGE: A
  high sensitivity particle tracker based on optically read out GEM}, Nucl.
  Instrum. Meth. A845 (2017) 285--288.
\newblock \href {https://doi.org/10.1016/j.nima.2016.04.014}
  {\path{doi:10.1016/j.nima.2016.04.014}}.

\bibitem{bib:jinst_orange2}
V.~C. Antochi, E.~Baracchini, G.~Cavoto, E.~D. Marco, M.~Marafini,
  G.~Mazzitelli, D.~Pinci, F.~Renga, S.~Tomassini, C.~Voena, {Combined readout
  of a triple-GEM detector}, JINST 13~(05) (2018) P05001.
\newblock \href {http://arxiv.org/abs/1803.06860} {\path{arXiv:1803.06860}},
  \href {https://doi.org/10.1088/1748-0221/13/05/P05001}
  {\path{doi:10.1088/1748-0221/13/05/P05001}}.

\bibitem{bib:btf}
G.~Mazzitelli, A.~Ghigo, F.~Sannibale, P.~Valente, G.~Vignola,
  \href{http://www.sciencedirect.com/science/article/pii/S0168900203023416}{Commissioning
  of the da$\phi$ne beam test facility}, Nuclear Instruments and Methods in
  Physics Research Section A: Accelerators, Spectrometers, Detectors and
  Associated Equipment 515~(3) (2003) 524 -- 542.
\newblock \href {https://doi.org/https://doi.org/10.1016/j.nima.2003.07.017}
  {\path{doi:https://doi.org/10.1016/j.nima.2003.07.017}}.
\newline\urlprefix\url{http://www.sciencedirect.com/science/article/pii/S0168900203023416}

\bibitem{Pinci:2019hhw}
I.~Abritta~Costa, et~al., {CYGNO: Triple-GEM Optical Readout for Directional
  Dark Matter Search}, J. Phys. Conf. Ser. 1498 (2020) 012016.
\newblock \href {http://arxiv.org/abs/1910.07277} {\path{arXiv:1910.07277}},
  \href {https://doi.org/10.1088/1742-6596/1498/1/012016}
  {\path{doi:10.1088/1742-6596/1498/1/012016}}.

\bibitem{bib:lewis}
P.~Lewis, S.~Vahsen, I.~Seong, M.~Hedges, I.~Jaegle, T.~Thorpe, {Absolute
  Position Measurement in a Gas Time Projection Chamber via Transverse
  Diffusion of Drift Charge}, Nucl. Instrum. Meth. A 789 (2015) 81--85.
\newblock \href {http://arxiv.org/abs/1410.1131} {\path{arXiv:1410.1131}},
  \href {https://doi.org/10.1016/j.nima.2015.03.024}
  {\path{doi:10.1016/j.nima.2015.03.024}}.

\bibitem{FENG201735}
D.~Feng, M.~Garcia-Sciveres, J.~Kadyk, A.~Wang,
  \href{http://www.sciencedirect.com/science/article/pii/S0168900217301523}{Determination
  of z coordinate from track width in minitpc}, Nuclear Instruments and Methods
  in Physics Research Section A: Accelerators, Spectrometers, Detectors and
  Associated Equipment 851 (2017) 35 -- 38.
\newblock \href {https://doi.org/https://doi.org/10.1016/j.nima.2017.01.061}
  {\path{doi:https://doi.org/10.1016/j.nima.2017.01.061}}.
\newline\urlprefix\url{http://www.sciencedirect.com/science/article/pii/S0168900217301523}

\bibitem{Battat:2015rna}
J.~Battat, et~al., {Reducing DRIFT Backgrounds with a Submicron
  Aluminized-Mylar Cathode}, Nucl. Instrum. Meth. A 794 (2015) 33--46.
\newblock \href {http://arxiv.org/abs/1502.03535} {\path{arXiv:1502.03535}},
  \href {https://doi.org/10.1016/j.nima.2015.04.070}
  {\path{doi:10.1016/j.nima.2015.04.070}}.

\bibitem{Daw:2013waa}
E.~Daw, et~al., {Long-term study of backgrounds in the DRIFT-II directional
  dark matter experiment}, JINST 9 (2014) P07021.
\newblock \href {http://arxiv.org/abs/1307.5525} {\path{arXiv:1307.5525}},
  \href {https://doi.org/10.1088/1748-0221/9/07/P07021}
  {\path{doi:10.1088/1748-0221/9/07/P07021}}.

\bibitem{bib:eps}
D.~Pinci, E.~Di~Marco, F.~Renga, C.~Voena, E.~Baracchini, G.~Mazzitelli,
  A.~Tomassini, G.~Cavoto, V.~C. Antochi, M.~Marafini, {Cygnus: development of
  a high resolution TPC for rare events}, PoS EPS-HEP2017 (2017) 077.
\newblock \href {https://doi.org/10.22323/1.314.0077}
  {\path{doi:10.22323/1.314.0077}}.

\bibitem{bib:ieee17}
G.~{Mazzitelli}, V.~C. {Antochi}, E.~{Baracchini}, G.~{Cavoto}, A.~{De Stena},
  E.~{Di Marco}, M.~{Marafini}, D.~{Pinci}, F.~{Renga}, S.~{Tomassini},
  C.~{Voena}, A high resolution tpc based on gem optical readout, in: 2017 IEEE
  Nuclear Science Symposium and Medical Imaging Conference (NSS/MIC), 2017, pp.
  1--4.
\newblock \href {https://doi.org/10.1109/NSSMIC.2017.8532631}
  {\path{doi:10.1109/NSSMIC.2017.8532631}}.

\bibitem{bib:elba}
D.~Pinci, E.~Baracchini, G.~Cavoto, E.~D. Marco, M.~Marafini, G.~Mazzitelli,
  F.~Renga, S.~Tomassini, C.~Voena,
  \href{http://www.sciencedirect.com/science/article/pii/S016890021831711X}{High
  resolution tpc based on optically readout gem}, Nuclear Instruments and
  Methods in Physics Research Section A: Accelerators, Spectrometers, Detectors
  and Associated Equipment (2018).
\newblock \href {https://doi.org/https://doi.org/10.1016/j.nima.2018.11.085}
  {\path{doi:https://doi.org/10.1016/j.nima.2018.11.085}}.
\newline\urlprefix\url{http://www.sciencedirect.com/science/article/pii/S016890021831711X}

\bibitem{Costa:2019tnu}
I.~A. Costa, E.~Baracchini, F.~Bellini, L.~Benussi, S.~Bianco, M.~Caponero,
  G.~Cavoto, G.~D'Imperio, E.~D. Marco, G.~Maccarrone, M.~Marafini,
  G.~Mazzitelli, A.~Messina, F.~Petrucci, D.~Piccolo, D.~Pinci, F.~Renga,
  F.~Rosatelli, G.~Saviano, S.~Tomassini,
  \href{https://doi.org/10.1088%2F1748-0221%2F14%2F07%2Fp07011}{Performance of
  optically readout {GEM}-based {TPC} with a 55fe source}, Journal of
  Instrumentation 14~(07) (2019) P07011--P07011.
\newblock \href {https://doi.org/10.1088/1748-0221/14/07/p07011}
  {\path{doi:10.1088/1748-0221/14/07/p07011}}.
\newline\urlprefix\url{https://doi.org/10.1088%2F1748-0221%2F14%2F07%2Fp07011}

\bibitem{baracchini2019cygno}
E.~Baracchini, R.~Bedogni, F.~Bellini, L.~Benussi, S.~Bianco, L.~Bignell,
  M.~Caponero, G.~Cavoto, E.~D. Marco, C.~Eldridge, A.~Ezeribe, R.~Gargana,
  T.~Gamble, R.~Gregorio, G.~Lane, D.~Loomba, W.~Lynch, G.~Maccarrone,
  M.~Marafini, G.~Mazzitelli, A.~Messina, A.~Mills, K.~Miuchi, F.~Petrucci,
  D.~Piccolo, D.~Pinci, N.~Phan, F.~Renga, G.~Saviano, N.~Spooner, T.~Thorpe,
  S.~Tomassini, S.~Vahsen, Cygno: a cygnus collaboration 1 m$^3$ module with
  optical readout for directional dark matter search (2019).
\newblock \href {http://arxiv.org/abs/1901.04190} {\path{arXiv:1901.04190}}.

\bibitem{Abe:2020bbf}
M.~Abe, et~al., {Development of a $\mu$-PIC with glass substrate aiming at high
  gas gain}, J. Phys. Conf. Ser. 1498~(1) (2020) 012002.
\newblock \href {https://doi.org/10.1088/1742-6596/1498/1/012002}
  {\path{doi:10.1088/1742-6596/1498/1/012002}}.

\bibitem{CYGNUSweb}
\url{https://web.infn.it/cygnus/}.

\bibitem{3dprinting}
\url{http://w3.lnf.infn.it/3d-printing-facility-lnf/}.

\bibitem{bib:stab}
E.~Baracchini, et~al., {Stability and detection performance of a GEM-based
  Optical Readout TPC with He/CF$_4$ gas mixtures}, JINST 15~(10) (2020)
  P10001.
\newblock \href {http://arxiv.org/abs/2007.00608} {\path{arXiv:2007.00608}},
  \href {https://doi.org/10.1088/1748-0221/15/10/P10001}
  {\path{doi:10.1088/1748-0221/15/10/P10001}}.

\bibitem{bib:thesis}
D.~Pinci, \href{http://weblib.cern.ch/abstract?CERN-THESIS-2006-070}{{A
  triple-GEM detector for the muon system of the LHCb experiment}}, Ph.D.
  thesis, Cagliari University, CERN-THESIS-2006-070 (2006).
\newline\urlprefix\url{http://weblib.cern.ch/abstract?CERN-THESIS-2006-070}

\bibitem{ORCAcamera}
\url{http://www.hamamatsu.com/jp/en/C13440-20CU.html}.

\bibitem{bib:cmos}
T.~Kugathasan, {Review on depleted CMOS}, PoS VERTEX2018 (2019) 042.
\newblock \href {https://doi.org/10.22323/1.348.0042}
  {\path{doi:10.22323/1.348.0042}}.

\bibitem{bib:ccd}
R.~C. Jared, T.~Y. Fujita, H.~G. Jackson, S.~B. Sidman, F.~S. Goulding, {Use of
  Ccd'$s$ in the Time Projection Chamber}, IEEE Trans. Nucl. Sci. 29 (1982)
  282.
\newblock \href {https://doi.org/10.1109/TNS.1982.4335846}
  {\path{doi:10.1109/TNS.1982.4335846}}.

\bibitem{bib:mesh}
F.~Kuger,
  \href{https://doi.org/10.1088%2F1748-0221%2F11%2F11%2Fc11043}{Micromesh-selection
  for the {ATLAS} new small wheel micromegas detectors}, Journal of
  Instrumentation 11~(11) (2016) C11043--C11043.
\newblock \href {https://doi.org/10.1088/1748-0221/11/11/c11043}
  {\path{doi:10.1088/1748-0221/11/11/c11043}}.
\newline\urlprefix\url{https://doi.org/10.1088%2F1748-0221%2F11%2F11%2Fc11043}

\bibitem{PMTPhotonics}
\url{http://www.hzcphotonics.com/products/XP3392.pdf}.

\bibitem{Kraus_2011}
V.~Kraus, M.~Holik, J.~Jakubek, M.~Kroupa, P.~Soukup, Z.~Vykydal,
  \href{https://doi.org/10.1088%2F1748-0221%2F6%2F01%2Fc01079}{{FITPix}
  {\textemdash} fast interface for timepix pixel detectors}, Journal of
  Instrumentation 6~(01) (2011) C01079.
\newblock \href {https://doi.org/10.1088/1748-0221/6/01/c01079}
  {\path{doi:10.1088/1748-0221/6/01/c01079}}.
\newline\urlprefix\url{https://doi.org/10.1088%2F1748-0221%2F6%2F01%2Fc01079}

\bibitem{Buonomo:2017sdz}
B.{Buonomo}, C.{Di Giulio}, L.G.{Foggetta}, P.{Valente}, A hardware and
  software overview on the new btf transverse profile monitor, in: Proc. 5th
  Int. Beam Instrumentation Conf. (IBIC'16), Barcelona, Spain, 2016, pp.
  818--821.
\newblock \href {https://doi.org/10.18429/JACoW-IBIC2016-WEPG73}
  {\path{doi:10.18429/JACoW-IBIC2016-WEPG73}}.

\bibitem{Buonomo:2017btf}
P.{Valente}, B.{Buonomo}, C.~{Di Giulio}, L.G.{Foggetta}, Frascati beam-test
  facility (btf) high resolution beam spot diagnostics, in: Proc. 5th Int. Beam
  Instrumentation Conf. (IBIC'16), Barcelona, Spain, 2016, pp. 222--225.
\newblock \href {https://doi.org/10.18429/JACoW-IBIC2016-MOPG65}
  {\path{doi:10.18429/JACoW-IBIC2016-MOPG65}}.

\bibitem{bib:daphne_time}
A.~Drago, G.~Di~Pirro, A.~Ghigo, F.~Sannibale, M.~Serio, {The DAPHNE timing
  system}, Conf. Proc. C 960610 (1996) 1775--1777.

\bibitem{FRAGA200388}
M.~Fraga, F.~Fraga, S.~Fetal, L.~Margato, R.~Marques, A.~Policarpo,
  \href{http://www.sciencedirect.com/science/article/pii/S0168900203007587}{The
  gem scintillation in he–cf4, ar–cf4, ar–tea and xe–tea mixtures},
  Nuclear Instruments and Methods in Physics Research Section A: Accelerators,
  Spectrometers, Detectors and Associated Equipment 504~(1) (2003) 88 -- 92,
  proceedings of the 3rd International Conference on New Developments in
  Photodetection.
\newblock \href {https://doi.org/https://doi.org/10.1016/S0168-9002(03)00758-7}
  {\path{doi:https://doi.org/10.1016/S0168-9002(03)00758-7}}.
\newline\urlprefix\url{http://www.sciencedirect.com/science/article/pii/S0168900203007587}

\bibitem{bib:roby}
R.~Campagnola,
  \href{https://cds.cern.ch/record/2313231/files/CERN-THESIS-2018-027.pdf}{{Study
  and optimization of the light-yield of a triple-GEM detector }}, Master's
  thesis, Sapienza University of Rome (2018).
\newline\urlprefix\url{https://cds.cern.ch/record/2313231/files/CERN-THESIS-2018-027.pdf}

\bibitem{bib:tesinatalia}
N.~Torchia, {Development of a tracker based on GEM optically readout },
  Master's thesis, Sapienza University of Rome (2016).

\bibitem{CAENN1570}
\url{http://www.caen.it/csite/CaenProd.jsp?parent=21\&idmod=894}.

\bibitem{Corradi:2007df}
G.~Corradi, F.~Murtas, D.~Tagnani, {A novel high-voltage system for a triple
  GEM detector}, Nucl. Instrum. Meth. A 572 (2007) 96--97.
\newblock \href {https://doi.org/10.1016/j.nima.2006.10.166}
  {\path{doi:10.1016/j.nima.2006.10.166}}.

\bibitem{bib:garfield}
R.~Veenhof, {GARFIELD, recent developments}, Nucl. Instrum. Meth. A 419 (1998)
  726--730.
\newblock \href {https://doi.org/10.1016/S0168-9002(98)00851-1}
  {\path{doi:10.1016/S0168-9002(98)00851-1}}.

\bibitem{bib:fe55}
I.~A. Costa, E.~Baracchini, F.~Bellini, L.~Benussi, S.~Bianco, M.~Caponero,
  G.~Cavoto, G.~D'Imperio, E.~D. Marco, G.~Maccarrone, M.~Marafini,
  G.~Mazzitelli, A.~Messina, F.~Petrucci, D.~Piccolo, D.~Pinci, F.~Renga,
  F.~Rosatelli, G.~Saviano, S.~Tomassini, Performance of optically readout
  {GEM}-based {TPC} with a 55fe source, Journal of Instrumentation 14~(07)
  (2019) P07011--P07011.
\newblock \href {https://doi.org/10.1088/1748-0221/14/07/p07011}
  {\path{doi:10.1088/1748-0221/14/07/p07011}}.

\bibitem{bib:hough}
R.~O. Duda, P.~E. Hart, \href{https://doi.org/10.1145/361237.361242}{Use of the
  hough transformation to detect lines and curves in pictures}, Commun. ACM
  15~(1) (1972) 11–15.
\newblock \href {https://doi.org/10.1145/361237.361242}
  {\path{doi:10.1145/361237.361242}}.
\newline\urlprefix\url{https://doi.org/10.1145/361237.361242}

\bibitem{bib:garfield1}
R.~Veenhof, {Garfield, a drift chamber simulation program}, Conf.\ Proc.\ C
  9306149 (1993) 66--71.

\bibitem{bib:2001mug}
V.~Golovatyuk, F.~Grancagnolo, R.~Perrino, {Influence of oxygen and moisture
  content on electron life time in helium\textendash{}isobutane gas mixtures},
  Nucl. Instrum. Meth. A 461~(1-3) (2001) 77--79.
\newblock \href {https://doi.org/10.1016/S0168-9002(00)01172-4}
  {\path{doi:10.1016/S0168-9002(00)01172-4}}.

\bibitem{bib:ieee_orange}
M.~Marafini, V.~Patera, D.~Pinci, A.~Sarti, A.~Sciubba, N.~M. Torchia, {Study
  of the Performance of an Optically Readout Triple-GEM}, IEEE Transactions on
  Nuclear Science 65 (2018) 604--608.
\newblock \href {https://doi.org/10.1109/TNS.2017.2778503}
  {\path{doi:10.1109/TNS.2017.2778503}}.

\bibitem{bib:rolandiblum}
W.~Blum, L.~Rolandi, W.~Riegler,
  \href{http://www.springer.com/physics/elementary/book/978-3-540-76683-4}{{Particle
  detection with drift chambers}}, Particle Acceleration and Detection, ISBN =
  9783540766834, 2008.
\newblock \href {https://doi.org/10.1007/978-3-540-76684-1}
  {\path{doi:10.1007/978-3-540-76684-1}}.
\newline\urlprefix\url{http://www.springer.com/physics/elementary/book/978-3-540-76683-4}

\bibitem{bib:clus}
E.~Baracchini, et~al., {A density-based clustering algorithm for the CYGNO data
  analysis}, JINST 15~(12) (2020) T12003.
\newblock \href {http://arxiv.org/abs/2007.01763} {\path{arXiv:2007.01763}},
  \href {https://doi.org/10.1088/1748-0221/15/12/T12003}
  {\path{doi:10.1088/1748-0221/15/12/T12003}}.

\bibitem{Battat:2014van}
J.~Battat, et~al., {First background-free limit from a directional dark matter
  experiment: results from a fully fiducialised DRIFT detector}, Phys.\ Dark
  Univ. 9-10 (2015) 1--7.
\newblock \href {http://arxiv.org/abs/1410.7821} {\path{arXiv:1410.7821}},
  \href {https://doi.org/10.1016/j.dark.2015.06.001}
  {\path{doi:10.1016/j.dark.2015.06.001}}.

\end{thebibliography}

\end{document}